\documentclass[12pt,preprint]{aastex}
\begin{document}
\newcommand{\etal}{et~al.}
\tighten
\title{Collimation, Proper Motions, and Physical Conditions in the HH 30 Jet
from HST Slitless Spectroscopy
\altaffilmark{1}}

\author{
	Patrick Hartigan \altaffilmark{2}, 
	and Jon Morse \altaffilmark{3,4},
	}

\vspace{1.0cm}

\altaffiltext{1}{Based on observations made with the NASA/ESA
{\it Hubble Space Telescope}, obtained at the Space Telescope Science
Institute, which is operated by the Association of Universities for
Research in Astronomy, Inc., under NASA contract NAS5-26555.}

\altaffiltext{2}{Rice University, Department of Physics and Astronomy,
6100 S. Main, Houston, TX 77521-1892, USA} 

\altaffiltext{3}{Arizona State University, Department of Physics and Astronomy,
P.O. Box 871504, Tempe, AZ 85287-1504, USA}

\altaffiltext{4}{Current address: NASA Goddard Space Flight Center, Code 665,
Greenbelt, MD, 20771}

\begin{abstract}

We present Space Telescope Imaging Spectrograph (STIS)
spectral images of the HH~30 stellar jet taken through
a wide slit over two epochs. The jet is unresolved spectrally, so the observations produce
emission-line images for each line in the spectrum. This rich dataset shows
how physical conditions in the jet vary with distance and time, produces precise proper
motions of knots within the jet, resolves the jet width close to the star, and
gives a spectrum of the reflected light from the disk over a large wavelength
range at several positions. 
We introduce a new method for analyzing a set of line ratios based on
minimizing a quadratic form between models and data. The method generates images
of the density, temperature and ionization fraction computed using all the 
possible line ratios appropriately weighted. In HH~30, the density declines
with distance from the source in a manner consistent with an expanding flow,
and is larger by a factor of two along the axis of the jet than it
is at the periphery. Ionization in the jet ranges from $\sim$ 5\%\ $-$ 40\%, and
high ionization/excitation knots form at about 100~AU from the star and propagate
outward with the flow. These high-excitation knots are not accompanied by
corresponding increases in the density, so if formed by velocity variations
the knots must have a strong internal magnetic pressure to smooth
out density increases while lengthening recombination times. 

\keywords{ISM: jets and outflows, Herbig-Haro objects; stars: pre-main-sequence; methods: data analysis}

\end{abstract}

\section{INTRODUCTION}

Included with the first catalog of Herbig-Haro objects \citep{lickbull}, 
HH~30 is in many ways the prototype for all jets from young stars.
HH~30 is located near the L1551 molecular cloud in the Taurus star formation
region at a distance of 140~pc \citep{kenyon94}.
\citet{mundt83} first noticed the jet-like nature of HH~30 from ground-based
images, and identified both a main jet and a counterjet. The system achieved
a great deal of fame when images from the Hubble Space Telescope (HST)
revealed a well-resolved, flared reflection
nebula on either side of an opaque circumstellar disk oriented nearly edge-on
\citep[][hereafter B96]{burrows96}, and a collimated clumpy jet emerging
exactly perpendicular to the disk plane on both sides of the disk.
The HST images made it clear that flared disks do not collimate high-velocity
optical jets on large scales;
instead, jets become collimated within a few tens of AU
of the disk plane, leading to the modern interpretation of magnetically-dominated
jets driven by accretion disks \citep[e.g.,][]{ferr06}.

The reflection nebulae at the top and bottom of the flared disk in HH~30
are bright, and quickly became the primary testbeds for models of
scattered light within accreting protostellar disks and envelopes
\citep{wood98a,cotera01,wood02}. The models show the disk inclination to
be 82 degrees, with the brighter (north) side of the flow tilted towards the Earth.
The nebulae vary significantly in brightness and in morphology,
reflecting conditions present at the base of the accretion disk where
material falls onto the star \citep[B96,][]{wood98b,wood00,watson04}.
Like other stellar jets, HH 30 extends to parsec scales
from the source \citep{lopez96}. The 
jet is not straight, but curves, possibly as a result
of precession (B96).  Seeing-limited ground-based observations 
typically do not resolve individual knots in HH~30 well. 
Jet widths inferred from deconvolved ground-based images 
($\sim$ 0.4$^{\prime\prime}$; \citealt{mundt91}) are about twice as large
as those found closer to the star with HST by \citet{ray96}, while
B96 report the jet to be unresolved everywhere even with HST. 

Because the HH 30 flow is in the plane of the sky and produces
a reliable source of new emission-line knots every few years, it is
an ideal target for studying jet physics. The nearly edge-on disk
blocks the stellar light effectively, so one can in principle trace the jet
to within a few tens of AU of the source and determine how physical
conditions such as the density, temperature, ionization fraction
and collimation behave as the jet first emerges from the accretion disk.
The orientation of the flow close to the plane of sky means there 
are no significant projection effects in images. More importantly, the internal
velocity dispersion is small \citep{app05}, and there are no strong bow shocks in
the flow, so the line widths in the jet are unresolved with typical
grating spectrometers.  Hence, a spectrum taken through a slit wide enough to include
the entire jet gives images of the jet in all the emission lines with a single
exposure. Moreover, one can remove the continuum light by subtracting background
at wavelengths adjacent to each emission line in the spectrum.

In this paper we present a set of deep (10 $-$ 30 ksec) images of
the HH~30 jet in all the emission lines visible in the jet 
from 3500\AA\ through 1$\mu$m. The new data are dithered, and
make it possible to follow the collimation in the jet down to $\sim$ 20~AU of the
source.  The images were taken at two epochs, so we can also
measure precise proper motions for the first time in this flow. We 
present the first high-quality spectrum of the reflection nebula from
the near-UV through the near-IR, and we quantify 
how colors vary with distance from the source in the reflection nebula.

But the main focus of this paper is on the
physical conditions in the jet, which we can only explore in
detail by having many line ratios to use as diagnostics.
Determining the best solution for the physical parameters of a plasma
from an observed set of non-independent line ratios is a common
problem in nebular astrophysics, and we discuss an analysis
technique that includes all the information present in the spectrum.
The resulting images of the ionization fraction, density, and temperature
of the jet at each position give new insights into how jets are 
heated as they emerge from the source.

\section{OBSERVATIONS AND DATA ANALYSIS}

Our program used the Space Telescope Imaging Spectrograph
(STIS) on HST to observe HH~30 in the
fall of 2000 (hereafter epoch 1) and again in the fall of 2002
(hereafter epoch 2; Table~1). The low-resolution gratings G430L 
and G750L were both used in epoch 1, and we reobserved HH~30 with
G430L in epoch 2.  The medium resolution grating G750M was used in
both epochs. The HH~30 relfection nebula was considerably brighter in the
first epoch than in the second epoch. The second epoch G430L observations are most
useful for quantifying how the reflection nebula changed with time,
while the two epochs with the G750M grating are ideal for
measuring proper motions in the resolved jet knots, which emit
brightly in the red lines of [O~I], [N~II], [S~II] and H$\alpha$.
The G430L observations were taken
over a time interval of about three weeks in the fall of 2000, but
the source showed no significant variability over this period so we
combined these observations into a single exposure. Likewise, the
G750M frames taken 14 Aug and 22 Aug 2002 are essentially identical,
and were coadded.

Standard pipeline processing is inadequate for these data because it leaves too many hot pixels in
the images and loses spatial resolution during the rectification process. We reduced
the data using tasks in the CALSTIS package of STSDAS within IRAF.
\footnote{ IRAF is distributed by the National Optical Astronomy Observatories,
    which are operated by the Association of Universities for Research
    in Astronomy, Inc., under cooperative agreement with the National
    Science Foundation. The Space Telescope Science Data Analysis System (STSDAS)
    is distributed by the Space Telescope Science Institute.}
A customized IRAF IMFORT task described by \citet{hep04} removed
hot pixels. Flux-calibrated 2D images have units
of surface brightness (erg cm$^{-2}$s$^{-1}$\AA $^{-1}$arcsec$^{-2}$), and we checked to
make sure that our flux calibrations matched those produced by pipeline reductions.
As a check of the flux calibration, the
surface brightness in epoch 1 of the brightest HH knot in the [O~I] 6300\AA\ G750L
spectrum agreed to better than 10\%\ with the G750M observations.

Each set of observations for
a given epoch and grating consists of at least four separate exposures dithered by fractional
pixels to improve the spatial resolution. The drizzle task within the STSDAS package
of IRAF combined individual images into composites \citep{fh02}.  The plate scale in the final 
composites is 0.02536 arcseconds per pixel, or 3.55 AU per pixel for a distance to
HH~30 of 140~pc. The spatial resolution of HST is close to the diffraction limit at
the wavelengths of the bright red emission lines in HH~30 ($\sim$ 0.07 arcseconds 
at 6500\AA).  After dithering, the G750L spectra have a dispersion of 2.441\AA\ per pixel
and cover 5234\AA\ -- 10230\AA, the G750M spectra have 0.277\AA\ per pixel between
6290\AA\ and 6857\AA , and the G430L spectra have 1.374\AA\ per pixel from
2907\AA\ to 5728\AA. The signal-to-noise-ratio (S/N)
in the reflection nebula is low blueward of $\sim$ 3200\AA\
and redward of $\sim$ 1$\mu$m, so we truncated the spectra accordingly.

The slit was 52 arcseconds in length and aligned at PA $-$147.35$^\circ$ to lie along the HH~30 jet. 
With the exception of one set of images taken through the 0.2 arcsecond
slit to check for internal velocity dispersion (see below), all data were taken through
the 2 arcsecond wide slit.
We performed a blind offset from a nearby star to center HH~30 in the slit. The closest star suitable for 
this offset was over 2 arcminutes away, which could introduce pointing errors of a few tenths of
an arcsecond. However, these errors are small compared with the size of the slit, and 
observations of the scattered light in the disk show that the slit was centered on the source
to within $\sim$ 0.2 arcseconds.
The C-shaped reflection nebula that
surrounds HH~30 extends about 2.5 arcseconds perpendicular to the slit, and
therefore fits almost entirely within the slit. Because the nebula also reflects
emission lines, an image of the reflection nebula appears near the base of the jet
for each of the strong emission lines. Figure 1 shows a portion of the G750M spectrum.

Observations of jets taken through slits that are wider than the jet produce emission-line
images at each line, provided the internal velocity dispersion of the emitting gas is 
unresolved. In our case the slit is $\sim$ 5 times wider than the jet (Fig.~1).
Internal velocity widths within HH jets are nearly always $\lesssim$ 100
km$\,$s$^{-1}$, and are unresolved with the low resolution gratings G430L and G750L.
However, we require the medium-resolution grating G750M
to image lines separated by 10 $-$ 20 \AA , such as 
[S~II]$\lambda$6716 and [S~II]$\lambda$6731, and [N~II]$\lambda$6583 and H$\alpha$.
With most jets, wide slit observations with G750M would be difficult or impossible
to interpret because spatial and spectral information are mixed. However, HH~30 lies nearly
in the plane of the sky \citep{wood00}, and has no strong bow shocks near the source that splatter material
along the line of sight. Hence, emission-line velocities and widths in the jet near the star
are low. \citet{app05} found that while scattered light generated a weak emission tail that
extends to $\sim$ 100 km$\,$s$^{-1}$ in the spectrum of the disk of HH~30, this emission is
down in flux by a factor of $\sim$ 100 from that of the jet, where the emission-line
profile lies entirely within a 40 km$\,$s$^{-1}$ interval. Coude spectra by \citet{mundt90}
also show narrow line widths in the jet close to the star.
An emission-line point source will be imaged over $\sim$ 2.5 pixels in the dithered
750M observations, corresponding to $\sim$ 0.7\AA , or $\sim$ 40 km$\,$s$^{-1}$,
so the jet knots should be unresolved in our spectra. To check this, we also imaged the jet
with the 0.2 arcsecond slit, and found the emission lines to be unresolved.  As a further
check, the G750L and G750M emission-line images of well-separated lines such as [O~I] 6300
and [O~I] 6363 are identical, which demonstrates that the jet is unresolved spectrally
in both gratings.

To create an image of a particular line, we simply extract a box centered on that line.
It is straightforward to measure the continuum from the reflection nebula on either side
of the emission line and subtract it from the image \citep[e.g.][]{hep04}. Images in various 
emission lines appear in Figure 2. The relative positions in the x-direction
(perpendicular to the jet) are in principle fixed by the wavelength solution. However,
the scatter in that solution of $\pm$ 0.3 pixels is improved by a factor of $\sim$ 2 by
shifting the x-positions
of the images to align the brightest knot in the jet. The jet knot used for alignment
appears fairly symmetrical in all the lines, so the uncertainty in its x-position is $\sim$ 0.1 pixel.

Aligning each emission-line image in the y-direction (along the jet) is more
involved because the star in HH~30 is not visible, no other stars lie along the slit,
and emission lines have offsets from one another in
the y-direction owing to heating and cooling along the flow. Fortunately, HH~30 has a bright
reflection nebula visible on either side of the disk, which appears as a dark lane in
images of the source \citep{watson04}, and it is easy to measure the position of this lane
in our spectral images.

At our request, K.~Wood and M.~O'Sullivan generated a series of 1D slices of a model of
the HH~30 reflected light at several wavelengths between 3000\AA\ and 9500\AA .  The
model slices are perpendicular to the disk through a 2-arcsecond slit, and show that the location
of the minimum of the reflection nebula at all wavelengths remains fixed to within a half pixel relative to
the projected location of the star. The models show that the star is located 0.092 $\pm$ 0.010
arcseconds away from the minimum of the reflected light. We measured the y-location of the
minimum of the reflection nebula adjacent to each emission line, and shifted each image
accordingly.

Spectra extracted from the flux-calibrated surface brightness images or ratios between
sections of such images must be corrected for the
point-spread function (PSF) of STIS, which worsens at the reddest
wavelengths. The calibration is accomplished with a photometric correction table. The values
affect the relative line ratios by $\lesssim$ 10\% for wavelengths $<$ 8500\AA , which includes
all of our lines except for the [C I] 9850 doublet, where the correction is $\sim$ 40\%.

\section{RESULTS}

\subsection{Reflection Nebula}

The reflection nebula at the base of the HH~30 jet has been known for
many years and has been the subject of several investigations. Most
recent studies make use of HST observations at
optical \citep[B96,][]{watson04} and near-infrared \citep{cotera01} wavelengths
to understand the scattering properties and spatial distribution of the dust.
The surface brightness of the nebula varies by as much as 1.5 magnitudes at optical
wavelengths, and there is evidence that reflected light from
hot spots on the stellar surface causes most of the observed asymmetries in
the nebula \citep{wood98b,watson04}. Further support for this idea comes from
observations of emission lines in the nebula, which have broad velocity tails
that are likely to be scattered light \citep{app05}.

The reflection nebula is easiest to study in the low-resolution (G430L and G750L) data
because the signal-to-noise in the continuum is higher than it is for the medium-resolution data. 
A single emission line reflected from the disk appears in the slitless data
as an image of the disk in that line, and occurs at some level in all the
bright lines in Figure 2, but is especially bright in the Balmer lines, which
come in part from the star.
Each row of the slitless images that crosses the disk (the jet lies along the columns) 
provides a spectrum of the reflected light that falls along that row, although
all absorption and emission lines will appear broadened by the physical size of the disk
in the extracted spectrum. 
One can obtain an integrated spectrum over a portion of the disk by simply summing the
appropriate range of rows.

Spectra of the entire disk show that the total flux of the reflected continuum varies
by over a magnitude at V, consistent with previously published results.
The blue spectra from 13 August 2000 and 3 September 2000 (Table~1) are nearly
identical, so we combined them into a single epoch. The nebula became fainter
in the second epoch of blue data (18 August 2002) by a factor that
increases steadily from about 2.0 at 3500\AA , to 3.0 at 5500\AA\ (Fig.~3).
A strong Balmer jump and many emission lines are present in the blue spectra.
The red spectrum taken 25 October 2000 was intermediate in flux between the 
two blue spectra in the region where the wavelengths overlap. 

The emission-line images in Fig.~2 show that the
broad base to the emission-line profiles in Fig.~3 come from the disk,
while the narrow cores arise in the jet. The ratio of the broad to narrow
components, measured from the medium-resolution red spectra (the lines
are not blended there), is about 0.9 for all the forbidden lines and $\sim$ 4 for
H$\alpha$. Both the forbidden lines and Balmer lines from the disk
are likely to be scattered light, with the star being the
primary source of illumination for the Balmer lines and the jet the
primary source for the forbidden lines.

The H$\alpha$ image of the disk in Fig.~4 shows an asymmetry that appears
in both epochs of the G750M data, with the H$\alpha$ brighter on the left (northwest).
The asymmetry persists in the G750L images, and so is not caused by spectrally resolved
scattered blueshifted emission from the outflow. The asymmetry is not present in the
forbidden lines, indicating that it originates in the strong H$\alpha$ emitting
region in the vicinity of the star.
\citet{watson04} noted similar asymmetries and attributed them to hot spots on the stellar
photosphere caused by accretion columns from the circumstellar disk.
The H$\alpha$ scattered light images appear truncated on the right (southeast), but
this position likely marks the edge of the slit, because this truncation
appears in the forbidden lines as well.

In addition to the Balmer jump and emission lines, there are weak TiO absorption
bands at 7800\AA\ and 7100\AA\ in the disk spectrum
\citep[Fig.~2, see also][]{app05}, but the signal-to-noise is
too low to constrain the spectral type of the star well. Using a STIS spectrum
of the M3.5 weak-line T Tauri star HBC 358 \citep{hk03} as a photospheric template,
the veiling is about 4 at these wavelengths, while the K7 wTTS Lk~Ca7 (primary) gives a
veiling of 1.5. By chance, the brightness of the nebula in epoch 1 agrees to within a few tenths
of a magnitude with the broadband V and I magnitudes reported by B96.

\citet{watson04} found that HH~30's reflection nebula became bluer further out in the
disk, and we also see this effect in our data. The ratio of the G430L spectrum of the outer
half ($\sim$ 1/3 arcsecond) of the blueshifted reflection nebula to a spectrum of the
inner half (closest to the star) rises in a nearly linear fashion from $\sim$ 1.0 at
5500\AA\ to $\sim$ 1.3 at 3500\AA .

\subsection{Proper Motions}

Knots in the blueshifted (northeastern) half of the HH~30 jet exhibited substantial proper motions of $\sim$ 15 pixels
in the 1.8 years between epoch 1 and epoch 2. To measure these motions
we combined the continuum-free images of the five brightest forbidden lines visible in the
G750M spectra $-$ [O~I] 6300, [O~I] 6363, [N~II] 6583, [S~II] 6716, and [S~II] 6731 $-$
into a single image for both epochs. We then used the code described in \citet{hartigan01}
to measure proper motions. The results appear in Table~2 and in Figure~5.
Velocities in the blueshifted jet are
rather low for an HH jet, $\sim$ 130 km$\,$s$^{-1}$. Velocity differences of $\sim$
30 km$\,$s$^{-1}$ are present, similar to those observed in other
HH flows \citep[e.g.,][]{hartigan01}. 

Velocities of knots in the redshifted counterjet are higher than they are in the main, blueshifted jet.
Three knots are present in the first epoch at distances of 0.63, 0.96, and 2.38 arcseconds from the star,
which we label D, E, and F, respectively. Knots E and F both have proper motions of $\sim$ 225 km$\,$s$^{-1}$.
The motion for knot D is less certain because knots exist in epoch 2 at distances 1.04 and 0.68
arcseconds from the star. Assuming the outer knot is knot D, the proper motion is $\sim$ 150 km$\,$s$^{-1}$.
It is possible that knot D had not emerged fully from the high extinction region of the disk plane
in epoch 1, which would result in a lower proper motion for this object. In any event, the redshifted 
portion of the flow is $\sim$ 1.7 times faster than its blueshifted counterpart.
Having one side of a flow much faster than the other side is not unusual for stellar jets
\citep[e.g., RW Aur;][]{lm03}.  However, in the case of HH~30 it does not appear that the knots
were ejected at the same time (Table 2).  There are knots in the HH~30 blueshifted jet 
at larger distances from the source than knot C in Fig.~5, but these
are either partially or completely excluded by our slit because the jet curves at
larger distances. The jet appears slightly tilted by about 2 degrees with respect to the 
position of the slit, and the motions of the knots all lie along the axis of the jet to
within the uncertainties of the observations.

It is instructive to compare our results with previous proper-motion measurements.
B96 determined proper motions in HH~30 with WFPC2 from broadband red images
separated by about a year. While HST can easily detect proper motions over this time
interval, the B96 measurements have large errors because the
first epoch exposure in particular was not very deep, and the images include 
bright continuum from the disk which makes it more difficult to measure accurate positions
for knots within an arcsecond of the star. 
Proper motions of the six knots in the blueshifted jet measured by B96 
have a large scatter, from as low as 54 km$\,$s$^{-1}$ to as high as 258 km$\,$s$^{-1}$,
with an average of 142 km$\,$s$^{-1}$. Ground-based images do not resolve individual knots well
close to the source, but \citet{mundt90} measured 170 $\pm$ 50 km$\,$s$^{-1}$
for the proper motion of a distant knot $\sim$ 10 arcsesconds from the star.

The combination of deep exposures (10 $-$ 30 ksec), a factor of four smaller pixel
scale in the dithered STIS images compared with WFPC2, precise continuum subtraction,
and improved analysis software greatly reduces the errors in the proper motions to
$\pm$ 4 km$\,$s$^{-1}$. With this precision it is possible to distinguish real velocity 
differences between adjacent knots and to predict where the
features we observe now should have appeared in the January 1995 images of B96
(the spatial resolution is too low in the ground-based images to make a useful comparison).
The knot labeled `A' in the blueshifted jet in Fig.~5 was ejected after 1995,
but knot `B' should have been located at a distance of 0.40 $\pm$ 0.03 arcseconds from the star
in January of 1995.  This position corresponds reasonably well, given the uncertainty
of the source position and the different filter used, with 
knot 9501-N, which B96 reported at a distance of 0.51 arcseconds. 
However, the B96 tangential velocity of only 54 km$\,$s$^{-1}$ is much lower than
our value of 149 $\pm$ 4 km$\,$s$^{-1}$ (Table 2). The difference may be caused
by uncertainties in measuring proper motions within the reflected light cavity of
the disk with the broadband B96 images.

Our object `C', which we measure as having
a proper motion of 125 $\pm$ 4 km$\,$s$^{-1}$, should have been located
2.11 $\pm$ 0.03 arcseconds from the star in January 1995. B96 report three
knots in this area, 9502-N (1.1 arcseconds, 258 km$\,$s$^{-1}$),
9503-N (2.0 arcseconds, 158 km$\,$s$^{-1}$), and 9504-N
(2.7 arcseconds, 84 km$\,$s$^{-1}$).  It appears that these three
objects visible in 1995 have now merged into a single structure --
object C extends for $\sim$ 1.8 arcseconds, about the distance between
9502-N and 9504-N in 1995. 

Some structural changes are evident in Fig.~5. Knot A became more elongated between
2000 and 2002, while knot B widened and faded, and the portion of knot C closest
to the star also faded somewhat. These changes are likely caused by internal
motions, shocks, rarefactions and cooling still unresolved in these images. However,
internal motions in these knots cannot exceed $\sim$ 40 km$\,$s$^{-1}$ to remain consistent
with observations of narrow intrinsic linewidths in these knots. A new knot (labeled `N' in
Figs.~6 and 7) was emerging from the source during our observations, but proper-motion
measurements are not possible because the feature has not yet detached clearly from the source.

\subsection{Jet Collimation}

The factor of four finer spatial scale of our dithered slitless STIS images relative
to narrowband WFPC2 images, the lack of continuum, better hot pixel removal strategies,
and long exposure times create an unprecedented database for measuring the width of the HH~30 jet.
The best images for this measurement are [S~II] 6716 and [S~II] 6731 with G750M, and
[O~I] 6300 with G750M and G750L, because in each of these cases the lines are bright and
unblended, and there is no stellar emission-line component.  We measured the FWHM by extracting spatial
profiles every three pixels along the jet in each emission-line image and fitting a Gaussian
to the profile with the SPLOT command in IRAF. Errors are dominated by uncertainties
in the baseline, and are larger near the star where the disk reflects the line emission in
the jet. The instrumental FWHM for slitless observations is that of the PSF of STIS. We
used the experimentally measured values and uncertainties of the PSF reported in
\citet{hep04} to deconvolve the FWHM, and propagated the errors of the measured FWHM with
the uncertainties in the deconvolution kernel to obtain errors for each deconvolved point. 

Results are shown in Fig.~6 for epoch 1, which have both G750L and G750M data,
and in Fig.~7 for epoch 2.  The widths measured from the G750L and G750M exposures in [O~I] 6300 
are identical, so the jet is unresolved spectrally in all images.
The upturn in the linewidth for the [O~I] points $\lesssim$ 15~AU is likely to be
instrumental in origin. The brightness of the jet drops sharply inside 20~AU, implying
less contrast with the reflected emission lines in the disk. Also, there will be some
light in the wings of the PSF from the bright jet at $\gtrsim$ 30~AU that contaminates
these measurements.

The collimation properties of the HH~30 jet have been the subject of
some debate in the literature. Using HST, B96 observed an opening angle of about 3 degrees, and
measured the width of the jet to within about 70 AU of the star. They concluded
that the jet was unresolved at this distance, with a FWHM $\lesssim$ 20~AU.
In contrast, \citet{ray96} analyzed the same image and found the jet to be spatially resolved
everywhere, with a FWHM $\sim$ 35~AU at a distance of 50~AU from the star, implying a very wide opening
angle for the jet at the source.  Deconvolving ground-based images with an FFT algorithm,
\citet{mundt91} reported an even larger FWHM of $\sim$ 60~AU at a distance of $\sim$ 350~AU, although
this measurement is very difficult to do because the jet is much narrower than the seeing disk
(which is $\sim$ 1 arcsecond or 140~AU FWHM). 

Our new observations settle this controversy.  Fig.~6 shows that the jet
is indeed spatially resolved everywhere by at least 5$\sigma$, though the width is only 
about half that reported by \citet{ray96} and by \citet{mundt91}, and is near the upper limit of B96.
The jet widens gradually from 14 $\pm$ 3~AU at 20~AU from the source to 36 $\pm$ 4~AU at 500~AU,
implying an opening half-angle of 2.6 $\pm$ 0.4 degrees, consistent with the opening angle
determined by B96 at larger distances.

The HH~30 jet is similar to other jets \citep[e.g.][]{hep04}
in that it does not project to a point at the source. However, this observation
does not mean that the jet emerges from a disk of radius 14~AU. Rather it is 
more likely that the opening angle of the flow is wider near the source, as
expected for a wind launched from an accretion disk.
There is no obvious correlation between the presence of
a bright knot and the width of the jet, as also noted by \cite{ray96}.
However, there is some structure in the
width of the jet, including relatively wide areas at about 230~AU and 310~AU during
epoch 1 (Fig.~6). These areas propagated downstream about 50~AU by epoch 2 (Fig.~7), at 
about the jet velocity. Apparently these sort of irregularities in the jet develop
near the source and the flow simply carries them along (see also Fig.~5).

On the fainter, redshifted side we coadded all the forbidden lines together to improve
the signal-to-noise but we were only able to measure jet widths within $\sim$ 400~AU 
of the star (Fig.~8).
The measurements are too uncertain to constrain the opening angle well, but the jet
is clearly resolved everywhere and is wider than it is on the blueshifted side.
At larger distances from the star, previous
images of the region \citep[][; B96]{mundt90,mundt91} have also shown
that the redshifted flow is wider and has a larger opening angle than its
blueshifted counterpart.
Being able to compare the opening angles, velocities, and widths in both a jet
and a counterjet is a potentially powerful diagnostic tool for analyzing the physics
of how jets are launched provided enough sources can be observed to derive meaningful statistics.

For a freely expanding flow, the observed opening half-angle
corresponds to a Mach number of 21.8. Using a jet velocity of 130 km$\,$s$^{-1}$
we find a sound speed of 6.0 km$\,$s$^{-1}$, which corresponds to a temperature of only
2600~K for a ratio of specific heats $\gamma$ = 5/3 and mean molecular weight 1.
This temperature is far too low to explain the fluxes in bright lines that originate
from upper levels such as [S~II] 4068.  The more likely value is $\sim$ 10 km$\,$s$^{-1}$,
which should occur for low-excitation forbidden lines in a mostly neutral
cooling zone of a shock. 

The observed opening angle is consistent with the temperature if there is
a source of confining pressure that is on the order of the thermal pressure.
The pressure cannot be provided by an ambient external medium, because 
shear between the jet and this medium would rapidly heat the interface between
the two fluids, creating strong shock waves in each that do not appear in the observations.
Instead, the most likely candidate for a confining pressure is a toroidal magnetic field.
Equating the magnetic pressure with the thermal pressure, using a total density
of $10^6$ cm$^{-3}$, and sound speed of 10 km$\,$s$^{-1}$ (T = 7260~K)
we find the magnetic field B $\sim$ 5 mG
at $\sim$ 300~AU, in agreement with the `fiducial' values of field strengths in
jets at this distance discussed by \citet{hartigan07}. 

\subsection{Emission Line Ratios within Individual Jet Knots}

Fig.~9 shows the observed ratio images of the four bright, red forbidden lines
for the two epochs. The signal-to-noise is excellent in the ratio images,
with uncertainties in the ratios of $<$ 5\%\ except within $\sim$ 50~AU of the source at the edges
of the jet, where reflected light introduces higher uncertainties. The
effect of scattered light is highest on the blue (left) side of the ratio images
that involve [N~II] 6583, because scattered
H$\alpha$ from the disk contaminates that side of the emission-line image.
The images were registered to the stellar position (marked with a horizontal
white line) as described in section 3.1.
The shorter horizontal lines in the figure mark locations
of bright knots N, A, B, and C in the summed forbidden line images for each epoch,
and are included as guides.  

The easiest emission-line ratio to interpret is that of [S~II] 6716/[S~II] 6731,
which decreases monotonically with increasing electron density. This ratio
image clearly shows that the electron density increases close to the source
\citep[see also][]{ber99}, although the ratio is in the high density limit close
to the star and therefore only gives a lower limit to the electron density in those areas.
The [S~II] ratio image also shows that the density declines near the edges of
the jet. This decline has been inferred from STIS mapping in other jets
\citep[e.g.][]{bacc00}, but Fig.~9
shows the effect with unprecedented spatial resolution and clarity.
The intensities in the combined forbidden line images in the two left panels
are far more concentrated along axis than is the electron density. 
Two effects contribute to this behavior:  first, the volume emissivity is
proportional to n$^2$ so small density fluctuations translate into larger
flux variations; and second, the brightness in the emission-line images
integrates along the line of sight, making the jet appear brighter along 
its axis even for a uniform density cylinder.

The second set of image ratios in Fig.~9 is [N~II] 6583/[S~II] 6716+6731.
This ratio depends primarily on the ionization fraction, but also has a weak positive
dependence on both the electron density and temperature. Hence,
this ratio follows `excitation' such as shocks, increasing wherever
the gas becomes hotter, denser, and more ionized. The N~II/S~II ratio images from
the two epochs are fascinating, and show substructure not apparent in the
individual images. The ratio is high in the 
two bright knots A and B, and these high excitation regions 
move outward with the flow velocity. The highest excitation portions of the
jet lie closer to the source than the brightest portions of the knots in both cases.
Additional structure in this ratio appears in the second epoch in the
vicinity of knot N.

There is a region of relatively low [N~II]/[S~II] near the source, also present in the
[N~II]$\lambda$6583/[O~I]$\lambda$6300 ratio (not shown). This decrease appears to be real and
not instrumental.  A registration error along the jet of 2 pixels between the two
images would produce a similar falloff of [N~II]/[S~II] near the source, but
the reflected light profiles in the original images show that registration
uncertainties are at most 0.5 pixels.  As a final check, the
[N~I]$\lambda$5199+5201/[N~II]$\lambda$6583 ratio also rises $<$ 30~AU from the source, indicating
that a drop in ionization fraction occurs there (see next section).
Beyond knot B, the [N~II]/[S~II] ratio declines as noted by \citet{ber99}. 

The final ratio shown in Fig.~9 is [O~I]$\lambda$6300/[S~II]$\lambda$6716+6731,
which is primarily sensitive to density, although the [O~I]$\lambda$6300 line also
has a positive dependence on temperature not present in the [S~II] lines
because the collision strength of [O~I] is a function of temperature. 
In a mostly neutral plasma like a stellar jet,
most oxygen is O~I, and essentially all sulfur is S~II. 
The upper levels for the [O~I]$\lambda$6300 and [S~II]$\lambda$6716+6731 transitions
have similar energies so the temperature dependence of the ratio is weak.
However, the critical density of [O~I]$\lambda$6300 ($\sim 10^6$ cm$^{-3}$)
is significantly higher than that for [S~II]$\lambda$6716+6731 ($\sim 10^4$ cm$^{-3}$),
so [O~I]$\lambda$6300 becomes stronger relative to [S~II]$\lambda$6716+6731 close to the star
where the density is high.  Comparing the [S~II]$\lambda$6716/[S~II]$\lambda$6731 ratio image
with [O~I]$\lambda$6300/[S~II]$\lambda$6716+6731 shows that the latter ratio increases sharply
as the S~II lines approach the high density limit, so together these three emission
lines constrain the density well everywhere in the jet. 

\section{Discussion}

\subsection{Physical Conditions Along the HH 30 Jet}

A main motivation of this work is to convert images of emission-line ratios
like those in Fig.~9 to images of T$_e$, N$_e$ and
X$_H$ = N$_{HII}$/(N$_{HII}$+N$_{HI}$) at each epoch. In this section we
describe how to create these images, consider how
these variables change along and perpendicular to the jet, and investigate
the sensitivity of results to assumptions such as abundance values
and charge exchange coupling. The results
give the first clear picture of how jets are heated as they
emerge from their sources.  Our data can also be used to infer the reddening
and mass loss rates but we defer discussion
of these issues to a second paper, where we also include blended
blue doublets such as [S~II]$\lambda$4068+4076 and [O~II]$\lambda$3727+3729 into the analysis.

\subsubsection{Determination of T$_e$, N$_e$, and X$_H$}

We use images of the four brightest forbidden lines obtained with the G750M setting,
[O~I]$\lambda$6300, [N~II]$\lambda$6583, [S~II]$\lambda$6716,
and [S~II]$\lambda$6731, to find T$_e$, N$_e$, and X$_H$
at each point in the jet where the S/N in the ratio between any pair of these lines is $\gtrsim$ 5. 
N~I, N~II, O~I, O~II, and S~II each have five levels populated in nebular conditions:
$^3$P, $^1$D, $^1$S for O~I and N~II and $^4$S, $^2$D, $^2$P  for O~II, N~I, and S~II.
We do not use [O~I]$\lambda$6363 or [N~II]$\lambda$6548 for anything other than flux calibration.
Ratios of [O~I]$\lambda$6363 / [O~I]$\lambda$6300 and
[N~II]$\lambda$6548 / [N~II]$\lambda$6583 are constant to within
the errors of measurement everywhere along the jet, and equal the ratios of their
respective A-values, as expected from lines that originate from a common upper level.
The other bright red line in the HH~30 jet is H$\alpha$, but this line has a prominent
reflected stellar component that contaminates the jet emission.

We constructed a model level population for N~I, N~II, O~I, O~II, and S~II for
a grid of N$_e$ and T$_e$ with the
atomic parameters summarized in Appendix B. With solar-like abundances (O=8.82, N=7.96, S=7.30
on a log scale with H=12.0) we calculated emission-line ratios for 
X$_H$ = 0 $-$ 1 in increments of 0.01, log N$_e$ = 2 $-$ 7 in steps of
0.1, and T$_e$/$10^4$K = 5 $-$ 25 in steps of 0.5 assuming the
ionization of N and O are tied to H by charge exchange as described below, and all S is S~II.
The latter assumption makes sense as no [S~III] or [S~I] lines are present in the spectrum. 
The best fit physical parameters were defined at each point
by minimizing the quadratic form C described in Appendix A over this
grid of 209100 models (100$\times$51$\times$41). 

The resulting images of N$_e$, T$_e$ and X$_H$ appear for each epoch
in Fig.~10, where, as in Fig.~9, the horizontal white line marks the location of the star
and short white lines denote the position of the knots in the jet at each epoch.
Dividing the electron density by X$_H$ produces a total density image,
and the product of T$_e$ and X$_H$ gives an `excitation' image that is
sensitive to hot ionized gas, as one should find behind a shock wave (Fig.~11).
The density images largely resemble what appears in the [S~II] image ratio in
Fig.~9, with the dominant behavior the decline of density with distance from
the source, as has been observed before at lower spatial resolution in the
1-D plots of \citet{ber99}. Density measurements are still possible
close to the source where the red [S~II] lines are in the high density
limit because the best fit solution also includes line ratios that involve
[O~I]$\lambda$6300, which is not in the high density limit.

The decline of the total density with distance shown in Fig.~12
resembles that produced by a radial conical flow that originates from
a finite source region. The jet is also a factor of two denser along
the axis than it is at its edges, and the bright portions of the jet
are typically the densest. The high-density knots clearly
move along the jet, and the higher S/N images of the second epoch show
what appear to be density filaments at the limit of the spatial resolution in
knot B. When interpreting these images one should keep in mind that the
observations integrate along the line of sight, which tends to smooth over
density gradients in the jet.

The images of X$_H$ and excitation show a rapid rise, followed by a
gradual decline with distance from the source \citep[see also][]{ber99}.
Figs.~11 and 12 show that this variation is not smooth; instead, the high excitation
regions show sharp, almost linear boundaries, and move outward with the flow.
The regions of high excitation occur on the side of the knot closest to the
source for both knots A and B. In knot A we see that this area of high
excitation appeared between the first and second epochs, indicating 
a heating event. 

The rise of the ionization/excitation in the first 100~AU
as the jet emerges from the source in a series of well-defined
heating events is a critical observational result of this paper 
that affects all models of jet launching, so it is important to verify
that the result does not depend upon the emission-line
ratios used in the analysis.  As noted above, using other lines
close to the source typically requires reddening corrections, but in this
case we can use the G430L [N~I]$\lambda$5199+5200 image to confirm the low value of
X$_H$ found close to the source from
the red line ratios. If the ionization fraction remained constant,
the [N~II]$\lambda$6583 / [N~I]$\lambda$5199+5201 ratio should increase steadily towards
the source because the critical density of [N~II]$\lambda$6583 exceeds that of
[N~I]$\lambda$5199+5201, and because the reddening increases toward the source.
However, images of this ratio in the last panel of Fig.~11 clearly show that
the ratio is {\it lower} close to the source, which can only happen if the
ionization fraction is lower there.

\subsubsection{Effects of Charge Exchange Coupling, Reddening, and Abundances}

The assumption that charge exchange sets the ionization fractions of O and N given the
ionization fraction of H is worth investigating further, as it is required
to solve for the hydrogen ionization fraction and temperature from a database that consists
of only [N~II]$\lambda$6583, [O~I]$\lambda$6300, [S~II]$\lambda$6716, and [S~II]$\lambda$6731. 
The charge exchange cross section between H and O is large
and has almost no temperature dependence because the ionization energies of H and O
are nearly the same. However, the charge exchange cross section is three orders of
magnitude lower for H-N than it is for H-O, so there
is some question as to how well the ionization
fractions of N are tied to those of H, especially since the ionization potentials of
H and N differ by $\sim$ 0.9eV.
\citet{be99} argue that the H-N charge exchange rate is high enough to fix the
(H~I/H~II)/(N~I/N~II) ratio as long as the hydrogen ionization fraction is $\lesssim$ 50\%,
but their model did not include photoionization, which affects H and N differently
owing to the different ionization potentials.

To test the validity of H-N charge exchange coupling we created a time-dependent
photoionization model, varying the ionizing spectrum and strength, and included
the atomic physics of collisional ionization, recombination, dielectronic recombination,
photoionization and charge exchange (see Appendix B for references) to determine
how closely the ionization fraction ratios of H and N follow those
predicted by charge exchange coupling.

The first case we tested was to singly ionize all the H, N, and O, and follow the
ionization fractions as the gas recombined. As expected, the (H~I/H~II)/(O~I/O~II)
ratio approached rapidly to the value predicted from charge exchange 
(to within 1 part in $10^4$ after 0.1 $\tau_{REC}(H)$, where 
$\tau_{REC}(H)$ is the H recombination timescale), and remained equal
to the charge exchange ratio as the gas recombined. The ionization fraction of
N was 10\%\ lower than the charge exchange value after 0.5 $\tau_{REC}(H)$
improving to 5\%\ lower after 2 $\tau_{REC}(H)$.  Experiments with input blackbody
photoionizing spectra yield similar results as long as the ionization fraction of
H is $\lesssim$ 90\%. Even the extreme case where the photoionizing flux is
low and H starts out completely ionized with N completely neutral (so that
charge exchange is solely responsible for ionizing N)
equilibrates to within $\sim$ 20\%\ of the charge exchange prediction
after one $\tau_{REC}(H)$. We conclude that the H-N charge exchange coupling 
assumption is well-justified.

Another issue is whether or not we should attempt to deredden the fluxes
of the four emission lines. In principle one might use the Balmer
decrement (H$\alpha$/H$\beta$) for this purpose.
The observed Balmer decrement in the bright knots varies between about
3.0 and 3.5, and rises to $\gtrsim$ 5 in the reflection
nebula and between the bright knots. This variation of the Balmer decrement
is likely to be caused by differing decrements 
in the scattered light from the star and the intrinsic emission from the jet.
Hence, we cannot use the decrement directly to find the reddening,
and instead must incorporate various blue doublets such
as [O~II]$\lambda$3727+3729, [S~II]$\lambda$4068+4076, and [N~I]$\lambda$5199+5201 into the analysis.
We defer this work to a subsequent paper. The Balmer decrement in
the bright jet knots indicates that A$_V$ $\lesssim$ 1, so
differential reddening between [O I]$\lambda$6300 and [S II]$\lambda$6731
is a factor of $\lesssim$ 1.04, and between [N~II]$\lambda$6583 and [S~II]$\lambda$6731
a factor of $\lesssim$ 1.02, small compared with the uncertainties in the relative
abundances of O, N, and S in the Taurus star forming region. 
We ignore these small effects of differential reddening for the current analysis. 

The density maps are largely independent of the abundances because the
[S~II]$\lambda$6716/[S~II]$\lambda$6731 ratio strongly constrains N$_e$ when log N$_e$ $\lesssim$ 4.
For a fixed set of observed emission-line ratios, reducing
the nitrogen abundance by a factor of two increases X$_H$ by a factor of 1.8 and
T$_e$ by a factor of 1.2. Overall, the fit to the line ratios becomes worse for
about 3/4 of the pixels and better for the remaining 1/4. Hence, it is possible
to constrain abundances with these data but the uncertainties are relatively high.
We also defer this analysis to a subsequent paper.

\section{Implications for Jet Heating}

The images presented in Figs. 9 $-$ 11 provide the best information
to date concerning how jets are heated as they emerge from their sources.
The observations show that some process heats and partially ionizes the jet
from $\lesssim$ 10\% within $\sim$ 20~AU of the star to
as high as 30\%\ -40\% at $\sim$ 100 AU from the star. The ionization
increase is not a steady process, but instead produces relatively
sharp boundaries, and areas of high ionization move outward with the
flow velocity.  The heating is not caused by photoionization from the
source, because the ionization fraction is initially low close to the source,
and increases outward. Similarly, a stationary focusing shock in the jet 
does not explain why the jet produces distinct knots with relatively high
ionization fractions, and there is no indication of any narrowing in the
jet that should accompany such a shock.

The most obvious candidates for heating and partially ionizing
the flow are shock waves produced by
supersonic variations in the flow velocity.  This scenario explains
why the jets are heated at some distance from the source, and internal
shocks produce sharp ionization features that propagate with the flow
velocity, in agreement with the observations.  However, one important
aspect of the data does not fit this model $-$ the density should
increase substantially in the postshock regions where the excitation
is high, and it does not.

It may be possible to produce sharp rises in the ionization fraction
without increasing the density substantially if the shocks are magnetic.
In this case the postshock pressure is dominated by the magnetic field,
and relaxes on the timescale for Alfven wave propagation, while 
the ionization in the shocked region relaxes on the recombination timescale.
Non-steady MHD jet models with atomic physics accurate
enough to follow ionization fractions have not yet been done, but 
the ionization signature of a supersonic magnetic collision should
linger substantially longer than the density enhancement does in stellar jets.
Magnetic field strengths in jets at distances of $\sim$ 100~AU from the source
are poorly understood, but recent models of time-dependent MHD pulses
indicate that the Alfven velocity could be $\gtrsim$ 50 km/s in these
regions \citep{hartigan07}, so an Alfven wave would cross the jet on a timescale of
$\lesssim$ 2 years. In comparison, the recombination time for
H gas with a density of $10^4$ cm$^{-3}$ is longer, $\sim$ 10 years,
approximately equal to the flow time from knot A to knot C.

The general decline of ionization
with distance means that more heating events occur close to the star
(as expected for stochastic velocity perturbations).
Beyond $\sim$ 1000~AU from the star, proper-motion measurements 
and high-resolution HST images of jets \citep[e.g.][]{hartigan01,heathcote96}
show that jets are heated by internal shock waves.
At these distances, small velocity perturbations have washed out, so
the jet gradually recombines and remains mostly neutral and cool until
it encounters a major working surface.

\section{SUMMARY}

We have constructed deep optical emission-line images of the HH~30 stellar jet
using STIS spectroscopy taken through a wide slit. The system has 
a fortunate geometry, with the axis of the jet nearly in the plane of the
sky. Hence, it is possible to observe the jet without projection effects
and we can trace the flow to within $\sim$ 20~AU of the star.
Low radial-velocity dispersions within the flow mean that the STIS
spectra are effectively slitless, and produce images for each emission line,
including those for which HST has no narrowband filters.  The resulting
data set, dithered for maximum spatial resolution and observed at two
epochs, enables us to diagnose physical conditions in the region where
the jet first emerges from the accretion disk.

Individual knots in the HH~30 jet show distinct proper motions between 
the 1.8 years that separate the two epochs. Motions in the blueshifted jet
range from 116 km$\,$s$^{-1}$ to 149 km$\,$s$^{-1}$, with uncertainties
of $\pm$ 4 km$\,$s$^{-1}$.  The jet has a resolved spatial width of
FWHM $\sim$ 14~AU at a distance of 20~AU from the source, and has a
constant opening half-angle of 2.6 degrees. The narrow opening angle
is consistent with the observed velocity and sound speed if there is
an additional confining magnetic pressure from toroidal fields in the jet.
Velocities in the redshifted counterjet
are $\sim$ 100 km$\,$s$^{-1}$ higher than they are in the main jet,
and the counterjet is less well-collimated than the main jet.

Spectra of the reflected light from the HH 30 disk reveal a veiled, 
late-type photosphere with both permitted and forbidden
emission lines and a prominent Balmer emission jump.
The reflected light from the disk varies substantially
in both morphology and in brightness, and is bluer at larger distances
from the star.

We developed a new analysis technique to find the best fit for
the electron temperature T$_e$, electron density N$_e$, and hydrogen ionization
fraction X$_H$ given a set of observed and model emission-line ratios.
The method, based on minimizing a quadratic form, has the advantage that
by construction it uses all the information present
in each available line ratio, and appropriately weights the fit of each ratio
by the uncertainty.

Our analysis focused on six ratios involving the four bright nebular lines 
of [N~II]$\lambda$6583, [O~I]$\lambda$6300, [S~II]$\lambda$6716, and [S~II]$\lambda$6731, with additional
ratios using [O~I]$\lambda$6363, [N~II]$\lambda$6548, H$\alpha$, H$\beta$, and
[N~I]$\lambda$5199+5200 used to verify the results. The images from each epoch
produce the first high-resolution images of T$_e$, N$_e$ and X$_H$
in a stellar jet. The density in the jet is highest close to the source,
and declines in a manner similar to that of radial conical flow from
a finite source region. The density in the jet is larger by a factor 
of two along the axis than at its edges. Distinct bright knots are also
denser, and propagate at the flow velocity. 

Maps of the ionization fraction and excitation
(defined as the product of T$_e$ and X$_H$)
show that the jet emerges with an ionization fraction of only $\sim$ 10\% ,
which increases to $\sim$ 30\%\ at $\sim$ 100~AU
from the source. Regions of locally higher ionization propagate with
the flow, and exhibit sharp spatial boundaries. At least one such
knot formed between the two epochs. Surprisingly, the high-excitation
knots are not accompanied by a density enhancement, suggesting that they may 
originate from velocity variability in a highly magnetized flow.

The two epochs reported in this paper were an accidental
benefit of a delay in HST scheduling. Stringent
ORIENT requirements limited visibility windows for the spacecraft,
and NICMOS took priority for all usable 2001 dates.
As a result, our data were taken
in two narrow time intervals, making it possible to 
study the motions and variability of the jet and the disk.
Stellar jets are dynamical systems, and we must watch them move to be
sure we understand the physics that governs them. Because of the
two epochs, we now know where the jet is heated and we are beginning to
understand the process that causes the heating.
Multiple epoch observations with the highest spatial resolution possible 
are clearly the best way to study these remarkable objects.

\vfill\eject
\section{APPENDIX A: Measuring Electron Densities, Electron Temperatures and Hydrogen Ionization Fractions
From Emission Line Ratios}

In this appendix we describe our method for extracting electron densities, temperatures,
and hydrogen ionization fractions from a set of emission-line ratios and uncertainties. 
In general, an observed emission-line ratio from an optically thin plasma where the dominant
processes are collisional excitation, collisional deexcitation and radiative decay
depends on the reddening, the relative abundances of the two elements, their respective
ionization fractions, the electron temperature and the electron density. 
One can use the flux ratio of lines with closely-spaced upper energy levels such as
[S~II]$\lambda$6716/[S~II]$\lambda$6731 to measure the electron density directly,
because these ratios are independent of abundances, reddening, and ionization fractions,
and insensitive to the electron temperature. This method has been applied to
HH objects for at least a half century now \citep{bohm56}. Other line ratios that arise from
the same ion but where the upper levels differ markedly in energy (e.g.
[O~I]$\lambda$5577/[O~I]$\lambda$6300]) depend on both the temperature and density, so with the density
in hand one can measure the temperature \citep{bbm81,hep04}. 

A variant of the above ideas introduced by \cite{be99} was to use the additional
constraint of strong charge exchange coupling between H and O, and H and N, which
ties the H~I/H~II ratio to O~I/O~II and to N~I/N~II given the temperature of the gas. 
Assuming standard abundance ratios of N/O, O/S and N/S, and taking all the S to be
S~II, one can then use [S~II]$\lambda$6716/[S~II]$\lambda$6731 to obtain the electron density as before,
and employ [N~II]$\lambda$6583/[O~I]$\lambda$6300,
[S~II]$\lambda$6716+6731/[O~I]$\lambda$6300, and [O~I]$\lambda$6300/[N~II]$\lambda$6583
to estimate the temperature and ionization fraction. Because these lines all have
similar upper level energies, the temperature determination is the most uncertain,
and relies mostly upon the temperature dependence of the collision strength of [O~I]
for the measurement. The main advantage of assuming that charge exchange ties the
ionization fraction of hydrogen to that of nitrogen and oxygen is that one can then
estimate the desired physical quantities from bright lines
in the less-extincted red part of the spectrum that are close enough in wavelength
to make differential reddening corrections negligible.

Estimates of N$_e$, T$_e$, and X$_H$ made in this manner implicitly assume that
these parameters do not vary substantially over a size scale equal
to the spatial resolution of the observations. However,
as shown clearly by \cite{bbm81}, and more recently by \cite{podio06}, temperatures
and densities of HH objects measured with different line ratios produce markedly
different results. The reason for this behavior is that HH objects are shock waves.
Emission lines from shocks radiate at different places in the cooling zones, 
with high-excitation lines and lower critical densities more prominent
near the front, and low-excitation, high critical density lines stronger at
larger distances.  Unless the cooling zones are resolved spatially, almost never the case with
ground based observations of HH jets,
constructing line ratios when the component lines sample different physical conditions can
lead to inconsistent results.  Hence, great care must be
taken when interpreting large scale trends in HH shocks when the cooling zones
are unresolved. Fortunately, HST resolves the cooling zones for most knots in jets
\citep{heathcote96}, so it is possible to apply the above analysis to many such images.

There are several aspects to the above methods of analysis that are less than ideal. The
[S~II]$\lambda$6716/[S~II]$\lambda$6731 ratio is usually in the high density limit close to the
source, and provides only a lower limit to the electron density there. Density
information is contained in other line ratios, but the analysis does not use
all the ratios. Moreover, it is unclear how to incorporate additional line fluxes into
the analysis. For example, if in addition to the four lines ([S~II]$\lambda$6716, [S~II]$\lambda$6731,
[N~II]$\lambda$6583, and [O~I]$\lambda$6300) used by \cite{be99} we also observe [N~I]$\lambda$5199+5201, what
is the best way to include this observation in the analysis? In principle, the new observation
defines four new emission-line ratios, each with a different uncertainty.

\subsection{Definition of the Problem}

{\it Given a set of n observed line fluxes F$_1$,F$_2$,...F$_n$, what is the best way to
determine the physical parameters N$_e$, T$_e$, and X$_H$ = n(H~II)/n(H~I+H~II)?}

We begin by discarding the flux normalization.
Absolute line flux measurements from ground-based spectra are rather rare, as they
require photometric conditions and are complicated by atmospheric dispersion and
pointing uncertainties. Physically,
the volume of emitting gas along the line of sight and the reddening determine
the absolute fluxes for these optically thin lines. In HH jets, which are inherently
clumpy, the absolute fluxes determine the volume filling factor of the emitting
material if one knows the width of the jet, but the absolute fluxes
do not affect estimates of T$_e$, N$_e$, and X$_H$, which depend only
upon the flux ratios. Hence, our analysis deals only with flux ratios.
The extra step of determining filling factors is easy to add after one
finds T$_e$, N$_e$, and X$_H$, assuming the spectra are flux-calibrated.

Most studies present line ratios relative to some normalization, typically
H$\beta$ or H$\alpha$ = 100. However, there is no reason {\it a-priori} to favor
any particular line for the normalization. Consider Table~3, which shows
artificial data for two models and an observation of three emission lines,
H$\alpha$, [O~I]$\lambda$6300 and [N~II]$\lambda$6583. At first glance it appears that model
B fits the observations better than model~A because the [N~II]
flux is off by a factor of two in both models, but model B gets the
[O~I] right while model A is again low by a factor of two. However, there
is an additional line ratio to consider, which is [O~I]/[N~II], and model~A
predicts this ratio correctly while model~B is off by a factor of two.
Hence, if we consider all line ratios equally, both model A and model~B
are equally good fits, because they both get one ratio correct and miss the
other ratios by factors of two.

\subsection{The Quadratic Form C}

Motivated by the above discussion, we seek a measure that will use {\it all} the 
emission-line observations in equal, and appropriately weighted manners to
determine the best fit set of the parameters N$_e$, T$_e$, and X$_H$. 
Given n line flux measurements we can construct p = n(n$-$1)/2 distinct pairs. The deviation
of the model from the data should be the same for a ratio and for its inverse,
and to ensure this behavior we define a new set of p ratios
r$_k$, where r$_k$ = ln(F$_i$/F$_j$) for some integers i, j = 1, 2, ..., n, with
i $<$ j.  The most obvious choice to determine a best fit model is to minimize
the quadratic form

$$
C = \sum_{k=1}^p {{\left(r_k - m_{k}\right)^{2}}\over{\sigma_k^2}}\eqno{(A1)}
$$

\noindent
where m$_{k}$ is the model prediction of the line ratio k from a given model,
and $\sigma_k$ is the observational uncertainty in the line ratio.

The quadratic form C has all the properties we desire for a line ratio fitting
algorithm. Each of the emission-line ratios contributes on an equal basis to
the value of C according to the uncertainty in the ratio. The best fit is
determined from line ratios and not fluxes, and information from each ratio
is automatically incorporated into the fit for the best physical parameters.
If, for example, ratio r$_i$ is unaffected by the temperature then it
will contribute equally to the value of C for all models that differ
only in the temperature. Line ratios that give only upper limits will
do so in the values of C as well. Adding new emission lines into the
analysis is trivial.  Applied to models A and B in Table~3, we
see that C$_A$ = C$_B$, as desired.  We define the `best' fit to a given
set of parameters T$_e$, N$_e$, and X$_H$ to be the one that minimizes C
for a set of observed line ratios.

\subsection{The Probability Distribution of C }

An outstanding issue is how one should interpret the values of C. 
Although this exercise has no bearing on the choice of the
best fit, which by definition is the one with the lowest value of C, 
it is still useful to investigate the
probability distribution of C because in a formal
sense the values of C determine confidence intervals for rejecting a null
hypothesis that the model exactly matches the data.

For astronomical spectra it is reasonable to take the individual
r$_k$ to be distributed normally.
Equation A1 looks a lot like a $\chi^2_n$ distribution, but it is not because
the r$_k$ are not independent of one another. For example, in Table~3 any two
line ratios suffice to determine the third. If {\bf m} and {\bf r} are the
p-dimensional vectors of a set of model ratios and observations, respectively,
and the r$_i$ have a normal distribution with mean m$_i$ and variance $\sigma_i^2$,
then a standard result from multivariate statistics is that the probability 
density function 

$$
{1\over{\left(2\pi\right)^{p/2}\sqrt{|{\bf V}|}}} exp\left[{-({\bf r}-{\bf m})^T
{\bf V}^{-1}({\bf r}-{\bf m}) }\over{2}\right] \eqno{(A2)}
$$

\noindent
is a multivariate normal, where {\bf V} is the covariance matrix of {\bf r}, and the superscript
`T' denotes a transpose.  Using the above notation, the quadratic form {\bf H} =
({\bf r}$-${\bf m})$^T${\bf V}$^{-1}$({\bf r}$-${\bf m}) is distributed as a
$\chi^2_n$ \citep{hc95}. In the case of independent r$_i$, the covariance matrix
is diagonal and we recover the standard definition of $\chi^2_n$.
While we could also minimize H to find the best fit, with the benefit that H
then has a known probability distribution, doing so requires inverting the
covariance matrix at each point within an emission-line map, an extra computational
step that did not seem warranted given that the quadratic form C exhibits all
the properties we desire for a best fit solution to the problem.

The probability distribution of C = {\bf x$^T$Ax} has been studied thoroughly in the mathematical
literature. Examples include independent normal variables \citep{cochran34}, normal
variables with nonzero covariance \citep{box54}, and arbitrarily distributed
vectors with nonzero covariance when {\bf A} = {\bf I} \citep{blacher03}. When,
as in our case, the x$_i$ are normally-distributed with nonzero covariance,
the probability distribution of C = $\sum_{i,j = 1}^p$A$_{ij}$x$_i$x$_j$
is given by a weighted sum of $\chi^2$ distributions, each with 1 degree of freedom:

$$
P(C) = \sum_{i=1}^p \lambda_i \chi^2_1 \eqno{(A3)}
$$

\noindent
where the weights $\lambda_i$ are the eigenvalues of {\bf VA}, with {\bf V}
the covariance matrix.  For our application, A$_{ij}$ = $\delta_{ij}\sigma_i^{-2}$.
Hence, the probability distribution differs for each set of line ratios depending on
the uncertainties in the ratios of the emission lines at that point.
For the simplest case of three line ratios with equal uncertainties $\sigma$
and equal covariances $\sigma_{ij}$, setting A$_{ij}$ = $\delta_{ij}\sigma^{-2}$,
b = $\sigma_{ij} / \sigma^2$ and keeping terms of order b (0 $<$ b $<$ 1)
in a power series expansion, we find the three eigenvalues $\lambda_i$ = 1, 
1$\pm\sqrt{3b}$.  The eigenvalues approach unity and the
distribution becomes $\chi^2_3$ as b$\rightarrow$0,
as expected for independent variables.

Experimenting with typical line ratios and uncertainties, the probability distribution
of C for p-ratios typically resembles a $\chi^2_{p}$ for large values of C, but the
actual distribution depends on the input data. For example, were we to include a very
low S/N emission line with a large uncertainty
into a list that initially has four emission lines, the four new
line ratios would not significantly increase the value of C even though p 
increases from 6 to 10. The best fit model would not change,
as desired, because the new noisy line does not affect C significantly.

\section{APPENDIX B: Atomic Parameters for Emission Line Models}

Energy levels for N~I, N~II, O~I, O~II, and S~II derive from the compilation
of \citet{m83}.  We used published values of the Einstein-A coefficients for the forbidden
transitions within N~I \citep{bz84}, N~II \citep{bell95,sz00,froese85},
O~I \citep{bz88}, O~II \citep{zeip82}, and S~II \citep{mz82}.

We interpolated collision strengths for N~I transitions 
from tabular values of \citet{bb81}, where we fit transitions to the $^4$S
level with log$_{10}(\Omega)$ = a + b*log$_{10}$(T/$10^4$) + c*log$_{10}$(T/$10^4$)$^2$,
with a, b, and c constants. Internal fine structure ratios in N~I were taken from
Table II of \citet{pradhan76}. Collision strengths within the $^2$P and $^2$D levels
used the power law fits of \citet{dmr76}. Collision strengths for the N~II lines
come from \citet{hb05}.  Collision strengths for O~I are from \citet{bb81}, except
for transitions within the $^3$P levels, which arise from power law fits we made to the
tabulated values of \citet{ln76}.  In a recent study of planetary nebulae, \citet{wang04}
found the O~II A-values of \citet{zeip82} to be more accurate than those of
\citet{zeip87}, and the collision strengths of \citet{pradhan76} to be better than
those of \citet{mb93}. Collision strengths for the lines of S~II are from 
\citet{keenan96}. 

Radiative recombination coefficients are from \citet{peq91}, and dielectronic
recombination coefficients are from \citet{ns83} at low temperatures and
\citet{lmf90} at high temperatures. We used the charge-exchange rates compiled by
\citet{kf96}, the collisional ionization rate coefficients of \citet{voronov97},
and the photoionization cross sections of \citet{verner96}.

\acknowledgements{We are grateful to the HST TAC for awarding enough time to complete this
project well. Thanks are due to K.~Wood and M.~O'Sullivan for sharing their scattered light
models of the reflection nebula in HH~30. This work has been supported under NASA/{\it HST}
grant GO-08238 from STScI, and NASA Origins grant NNG05GH97G.}

\begin{center}
\begin{deluxetable}{lccc}
\singlespace
\tablenum{1}
\tablewidth{0pt}
\tablecolumns{4}
\tabcolsep = 0.08in
\parindent=0em
\tablecaption{STIS Observing Log}
\tablehead{
\colhead{Grating} & \colhead{Number of} & \colhead{Total Exposure} &  \colhead{Dates}\\
\colhead{}        & \colhead{Exposures$^a$} & \colhead{Time (sec)} &  \colhead{}}
\startdata
G430L& 5&13323&13 Aug \& 3 Sept 2000\\
G430L& 4&10538&18 Aug 2002\\
G750M$^b$& 1&2146&3 Sep 2000\\
G750M& 4&10467&25 Oct 2000\\
G750M&12&31404&14 Aug \& 22 Aug 2002\\
G750L& 4&10262&25 Oct 2000\\
\enddata
\tablenotetext{a} {Each exposure was CR-split, and used the 52$\times$2 slit except as noted.}
\tablenotetext{b} {52$\times$0.2 slit.}
\end{deluxetable}
\end{center}

\begin{center}
\begin{deluxetable}{cccccc}
\singlespace
\tablenum{2}
\tablewidth{0pt}
\tablecolumns{6}
\tabcolsep = 0.08in
\parindent=0em
\tablecaption{Proper Motions in the HH 30 Jet}
\tablehead{
\colhead{Object$^a$} & \colhead{Separation$^b$} & \colhead{$\Delta$ X$^c$} & \colhead{$\Delta$ Y$^c$} &  \colhead{Velocity$^d$}& \colhead{Age$^e$}\\
}
\startdata
A&0.78&  1 (2)& -175 (5)  &116 (4)&4.5\\
B&1.70& -1 (3)& -224 (5)  &149 (4)&7.6\\
C&3.20& -7 (2)& -188 (5)  &125 (4)&17.0\\
D&-0.63& 19 (5)& -227 (5) &151 (5)&2.8\\
E&-0.96& 31 (10)& -325 (10)&217 (9)&3.0\\
F&-3.20& 13 (9)& -343 (10) &229 (9)&6.9\\
\enddata
\tablenotetext{a} {Object boundaries are defined in Figure 5.}
\tablenotetext{b} {Distance in arcseconds from the center of the box that defines the object to the HH 30 star for epoch 1. A negative number corresponds to the redshifted counterjet.}
\tablenotetext{c} {Proper motions (and errors) in milli-arcseconds per year. 
The X and Y-axes are defined as in Figure 5, with the negative Y-axis corresponding to motion along the
jet for the blueshifted portion of the flow.}
\tablenotetext{d} {Velocity in the plane of the sky in km$\,$s$^{-1}$ for a distance of 140~pc.}
\tablenotetext{e} {Age of the knot in years relative to 2000.82~UT, the first epoch of observations.}
\end{deluxetable}
\end{center}

\begin{center}
\begin{deluxetable}{cccc}
\singlespace
\tablenum{3}
\tablewidth{0pt}
\tablecolumns{4}
\tabcolsep = 0.08in
\parindent=0em
\tablecaption{Example of Observed and Model Fluxes for Three Lines}
\tablehead{
\colhead{Line} & \colhead{Observation} & \colhead{Model A} & \colhead{Model B}
}
\startdata
H$\alpha$     & 1.00 (0.10) & 1.00 & 1.00\\
{[O I]} 6300  & 0.40 (0.04) & 0.20 & 0.40\\
{[N II]} 6583 & 0.20 (0.02) & 0.10 & 0.40\\
\\
\enddata
\end{deluxetable}
\end{center}

\null\vfill\eject

\pagestyle{empty}

\begin{figure}
\def\thefigure{1}
\plotone{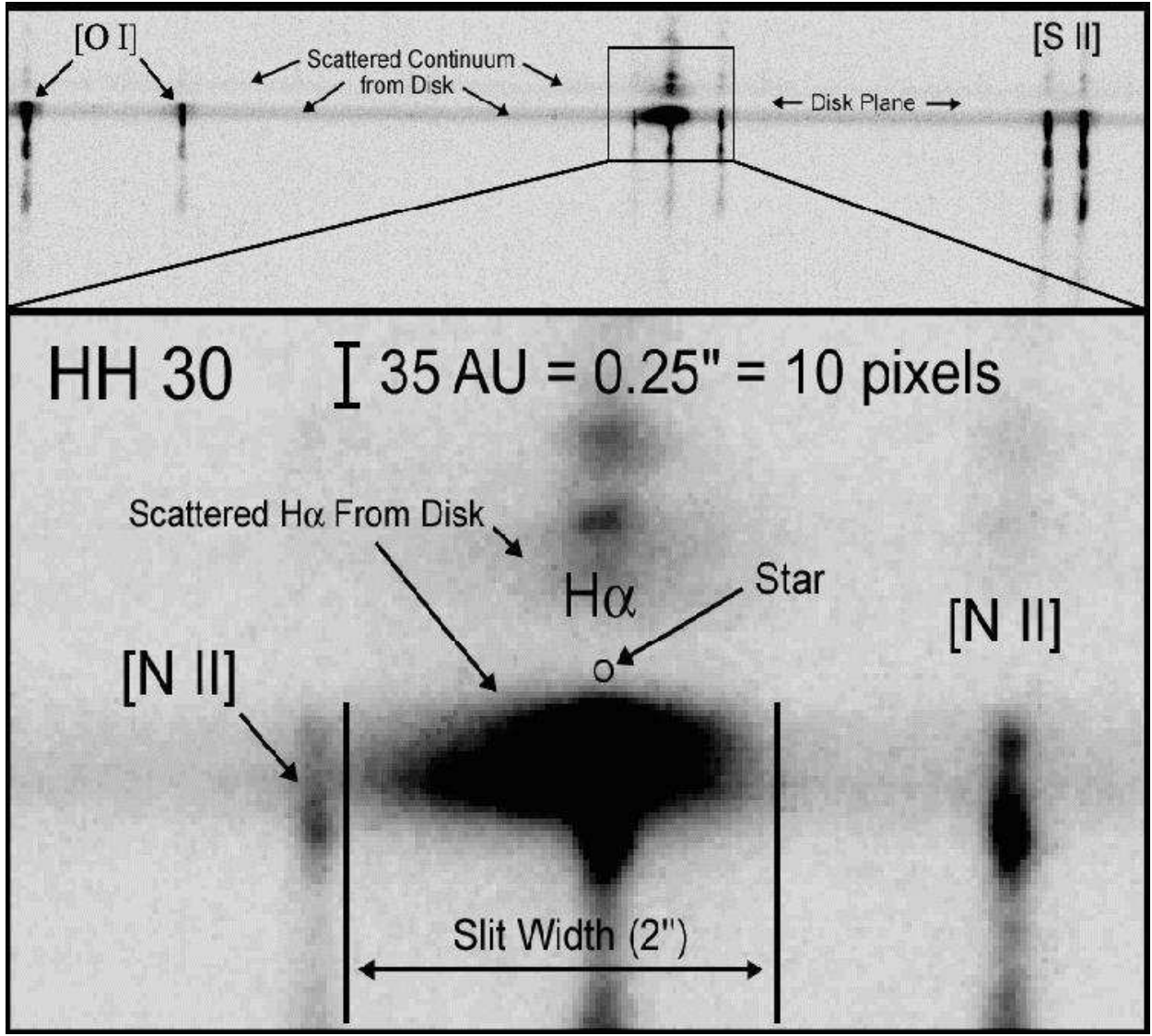}
\caption{Spectral images of the HH~30 system with the G750M grating. The locations of the star,
disk plane, and bright emission lines are marked. Reflected light is present at the top and the
bottom of the opaque disk, which is nearly edge-on in this object. The brightest (blueshifted)
portion of the jet extends below the disk (to the northeast) in this and subsequent figures,
while a fainter redshifted counterjet exists above the disk plane (to the southwest).
The slit is much wider than the jet, so these data generate an image for each emission line.}
\end{figure}

\begin{figure}
\def\thefigure{2}
\vbox to 5.5in{\null\vfill\includegraphics{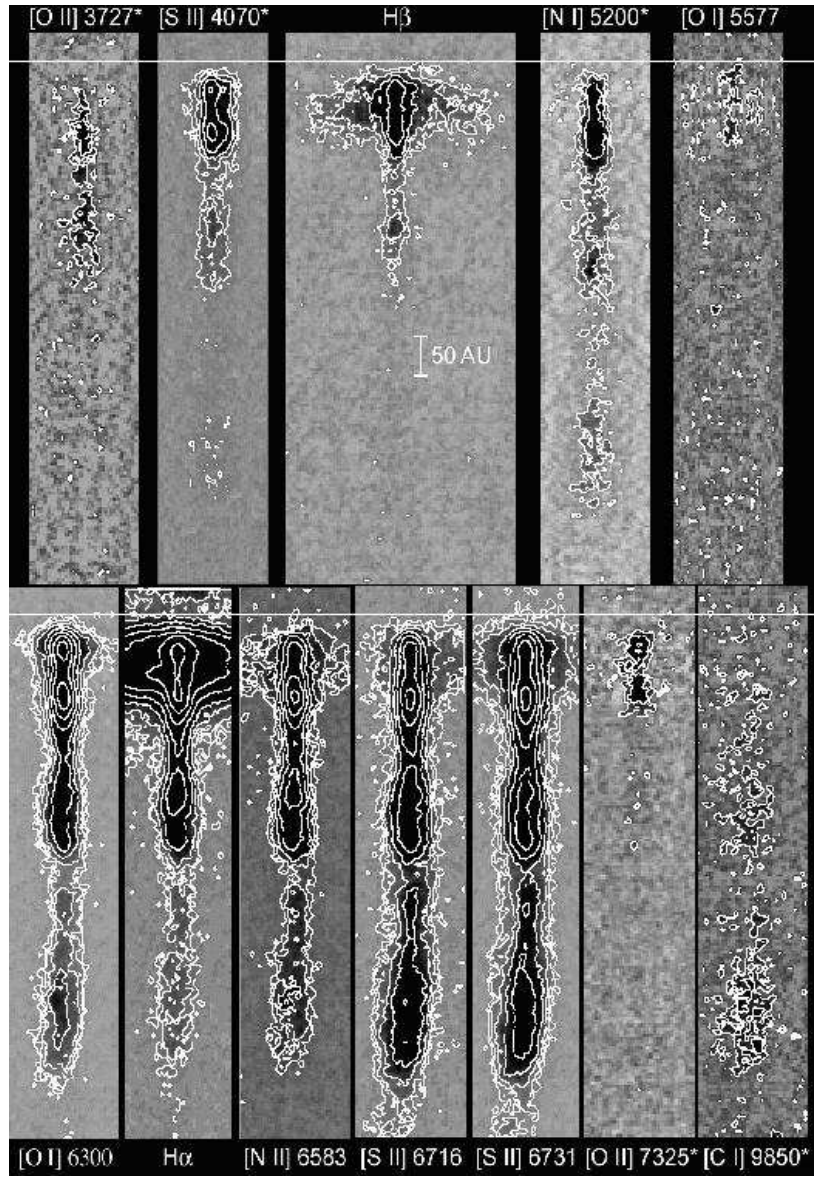}}
\caption{Epoch 1 emission-line images of the HH 30 jet. The slit was oriented at a position angle
of $-$147 degrees. Images of [O~II]$\lambda$3727, [S~II]$\lambda$4070, H$\beta$,
[N~I]$\lambda$5200, and [O~I]$\lambda$5577 are from G430L, while images of
[O~I]$\lambda$6300, H$\alpha$, [N~II]$\lambda$6583, [S~II]$\lambda$6716, and
[S~II]$\lambda$6731 are from G750M, and images of [O~II]$\lambda$7325 and
[C~I]$\lambda$9850 are from G750L. Adjacent contours are spaced by a
factor of two, with the lowest contour level (in units of $10^{-15}$
erg$\,$cm$^{-2}$s$^{-1}$arcsec$^{-2}$) equal to 3.47 for [O~II]$\lambda$3727 and
[S~II]$\lambda$4070, 2.17 for H$\beta$ and [O~I]$\lambda$5577, 1.73 for [N~I]$\lambda$5200,
1.93 for [O~II]$\lambda$7325, 3.08 for [C~I]$\lambda$9850, and 0.70 for all the other
lines. Line identifications labeled with an asterisk are doublets or quartets
that are partially resolved spectrally. The horizontal white line marks the location of the source.
}
\end{figure}

\begin{figure}
\def\thefigure{3}
\null\vbox to 0.3in{ }
\vbox to 5.5in{\null\vfill\includegraphics{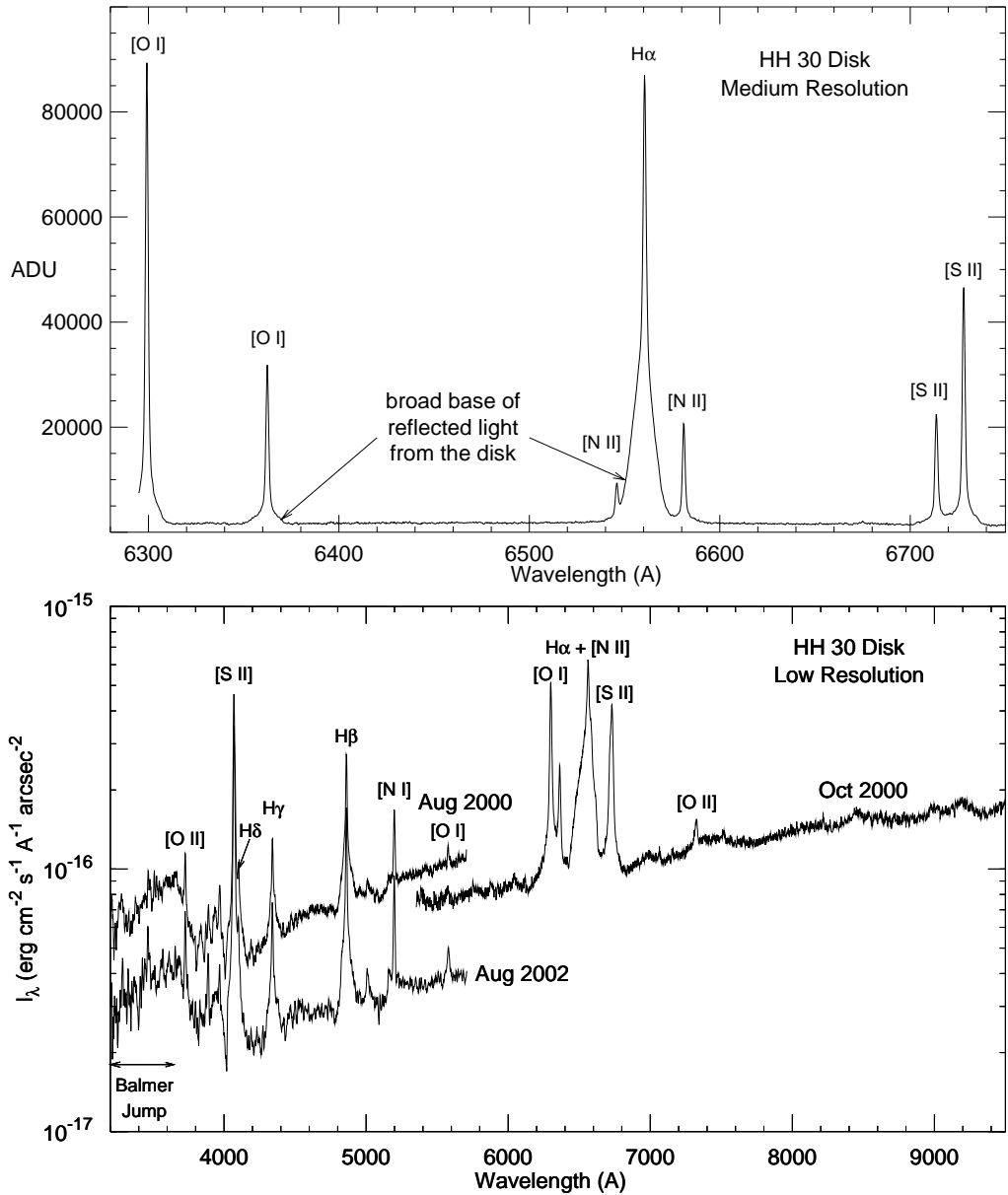}}
\caption{ Spectra of the HH~30 Disk. Top: A medium-resolution spectrum taken 14 Aug 2002, obtained by
integrating the entire light of the disk and any emission from the jet that is superposed atop the
disk. The broad base to the profiles results from emission lines reflected from the disk,
and not from high-velocity material. The narrow emission comes from spatially confined knots in the jet.
Bottom: Low-resolution spectra of the entire HH 30 disk. Vertical offsets between the spectra are caused by
flux variability between the observations. Weak TiO
bands are present in the red spectrum around 7100\AA\ and 7800\AA .}
\end{figure}

\begin{figure}
\def\thefigure{4}
\plotone{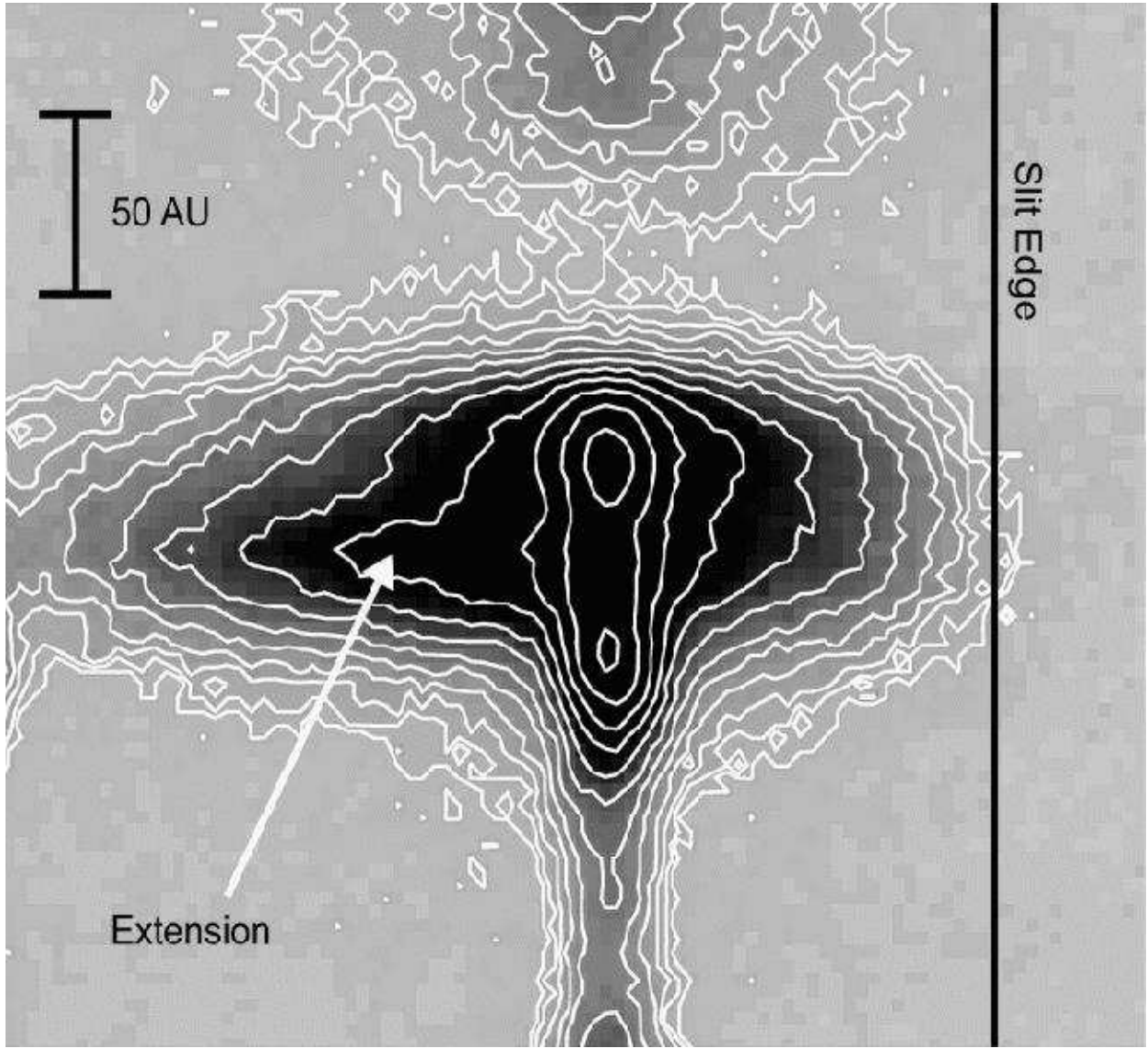}
\caption{An H$\alpha$ image of the HH~30 disk and jet taken with the G750M grating on 25 Oct, 2000.
The disk appears in scattered light, and has a bright extension to the left. 
The slit partially truncates the disk on the right side.
Adjacent contours are separated by a factor of $\sqrt{2}$, and the lowest contour
is $2.2\times 10^{-15}$ erg cm$^{-2}$s$^{-1}$arcsec$^{-2}$.
}
\end{figure}
\vfill\eject

\begin{figure}
\def\thefigure{5}
\vbox to 8.35in{\null\vfill\includegraphics{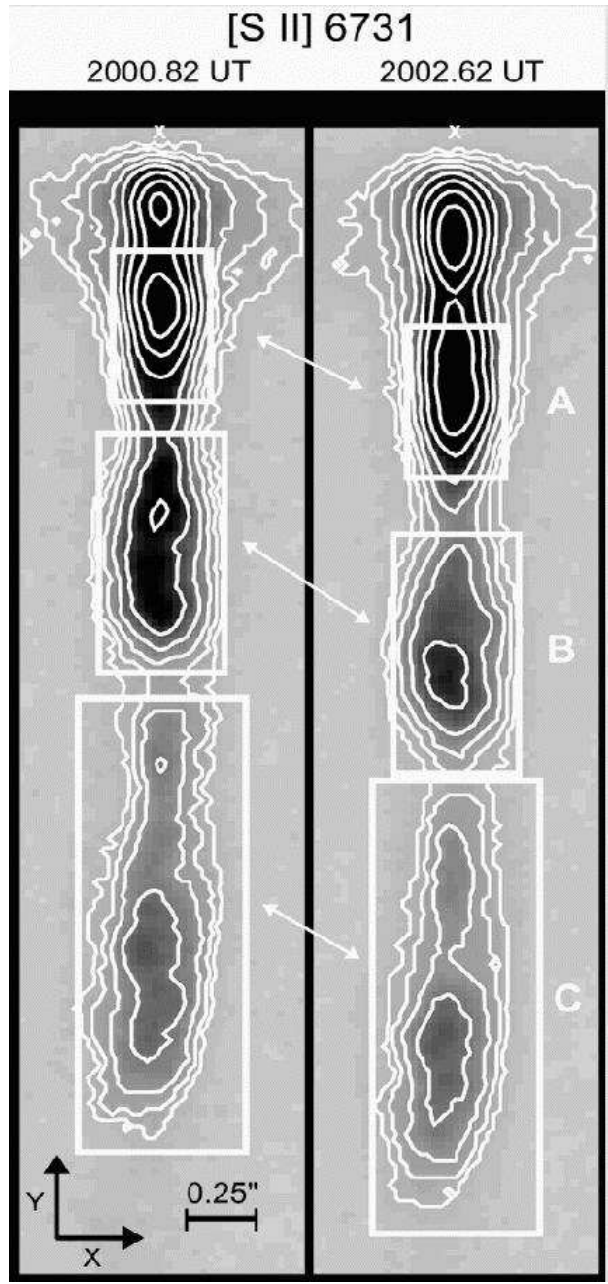}
\caption{Proper motions in the blueshifted component of the HH 30 jet.
The objects labeled A, B, and C move $\sim$ 0.4 arcseconds (56~AU)
over a time interval of 1.8 years. Contours are spaced by a factor of $\sqrt{3}$ in these [S~II]
images, with the lowest contour $1.1\times 10^{-15}$ erg cm$^{-2}$s$^{-1}$arcsec$^{-2}$. The source
position is denoted with an `X'. A new knot is also emerging from the source.}
}
\end{figure}

\begin{figure}
\def\thefigure{6}
\vbox to 8.35in{\null\vfill\includegraphics{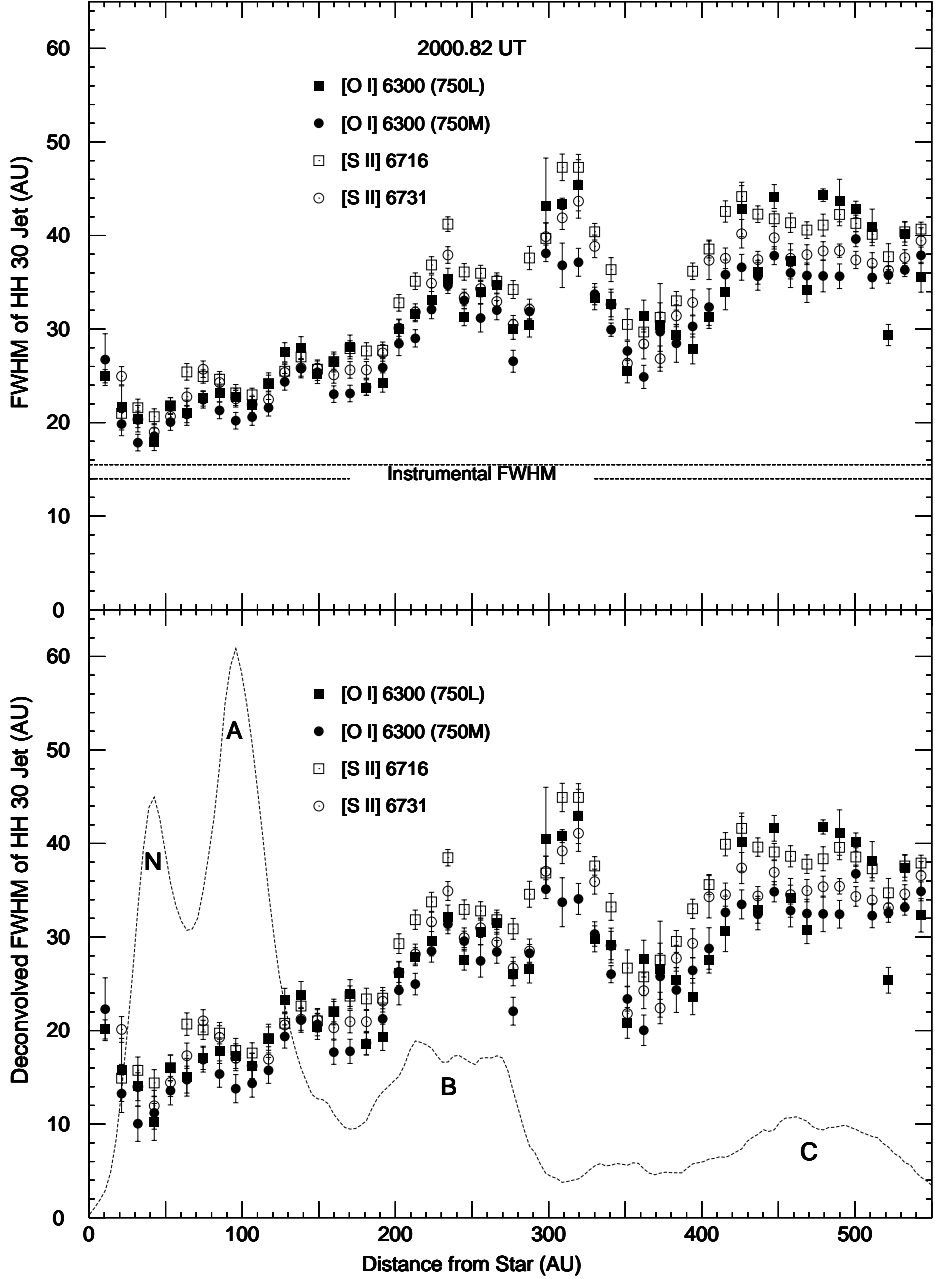}
\caption{Top: Measured FWHM of the jet as a function of distance from the source in four emission
line images taken during epoch 1. One arcsecond corresponds to 140~AU. The two horizontal lines
mark the upper and lower limits to the measured instrumental FWHM. Each point represents an average of
three lines. Bottom: Deconvolved jet widths. The jet width gradually increases with distance from the
source. The curve is the relative intensity of the forbidden lines with distance, obtained by
adding together all the bright red emission-line images and coadding the fluxes for each row.
Knots labeled A, B, and C are used to measure proper motions. A new knot `N' is emerging from
the source. Extinction from the nearly edge-on disk obscures the jet at distances $\lesssim$ 20~AU
from the source.  There is no correlation of the FWHM with the presence or absence of a bright emission knot.}
}
\end{figure}

\begin{figure}
\def\thefigure{7}
\vbox to 8.35in{\null\vfill\includegraphics{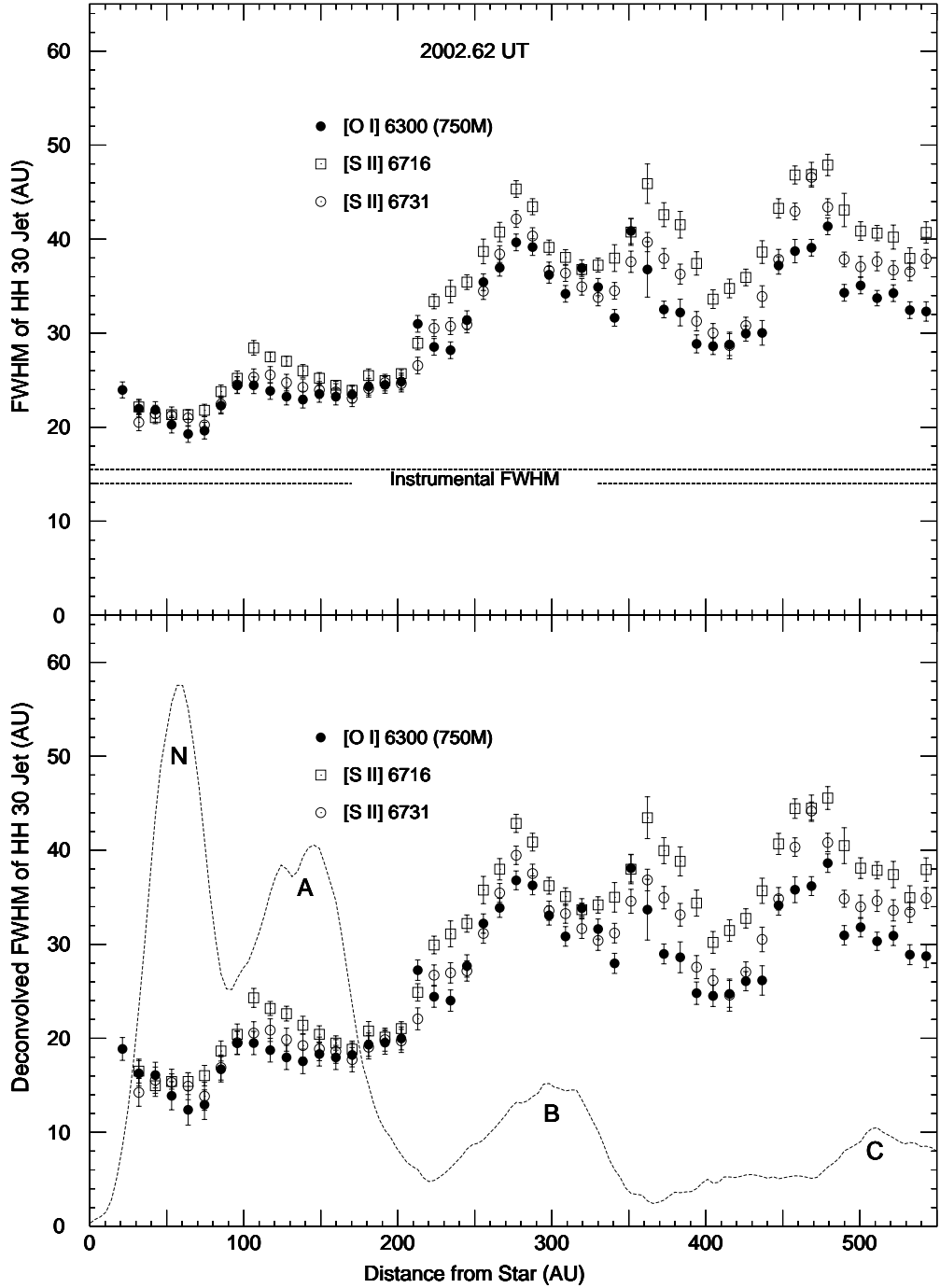}
\caption{Same as Fig.~6 but for epoch 2.
Irregularities in the jet width have propagated outward with the jet velocity since the
first epoch in Fig.~6.}
}
\end{figure}

\begin{figure}
\def\thefigure{8}
\vbox to 8.35in{\null\vfill\includegraphics{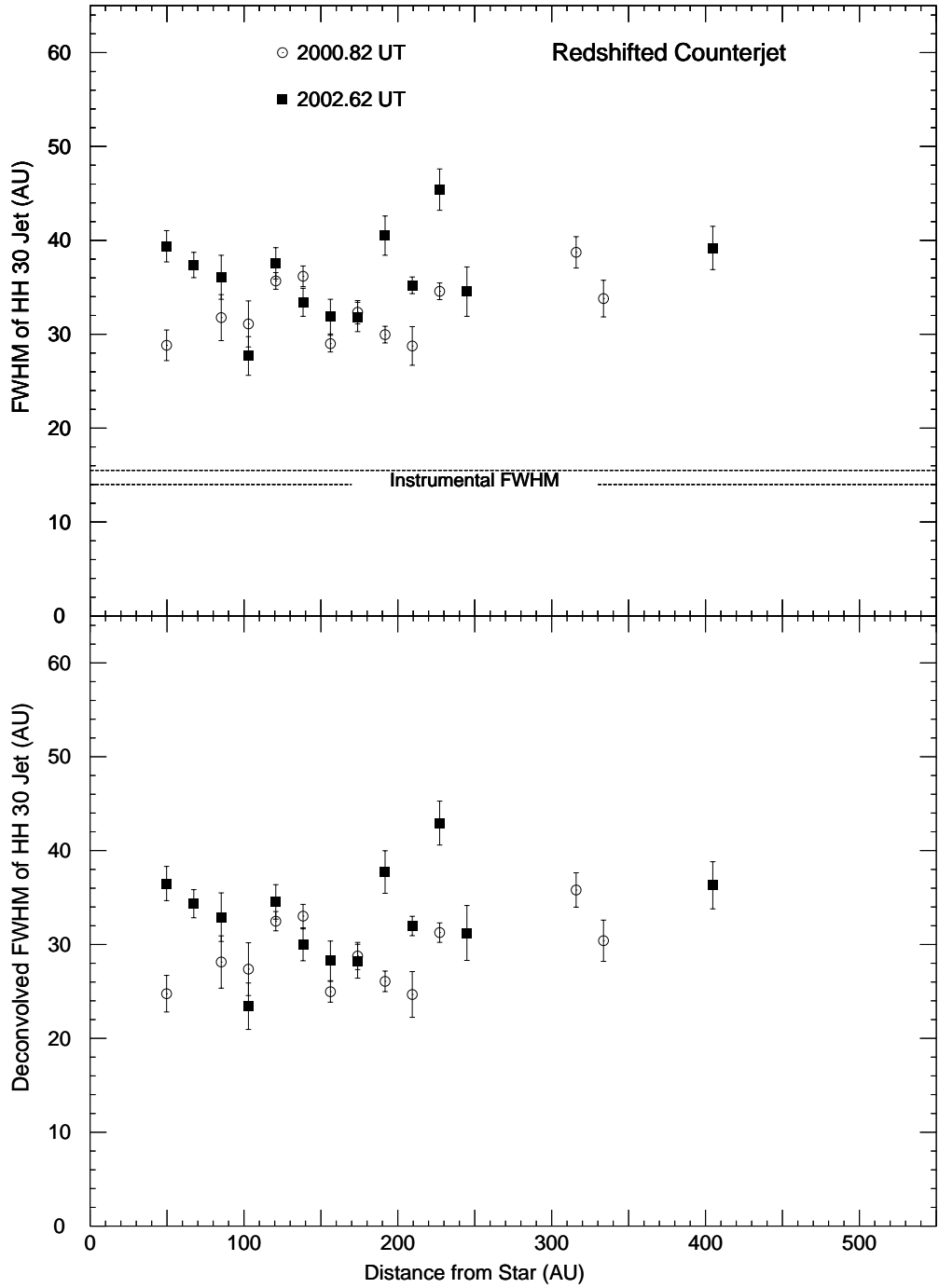}
\caption{Same as Fig.~6 but for the redshifted counterjet using
the sum of [O~I] 6300, [O~I] 6363, [N~II] 6583, [S~II] 6716
and [S~II] 6731. The Gaussian FWHM for each point was measured from a spatial cut perpendicular
to the jet, coadding five rows (0.125 arcseconds) for each point.
The jet is wider on the redshifted side than it is on the blueshifted side.}
}
\end{figure}

\begin{figure}
\def\thefigure{9}
\vbox to 8.35in{\null\vfill\includegraphics{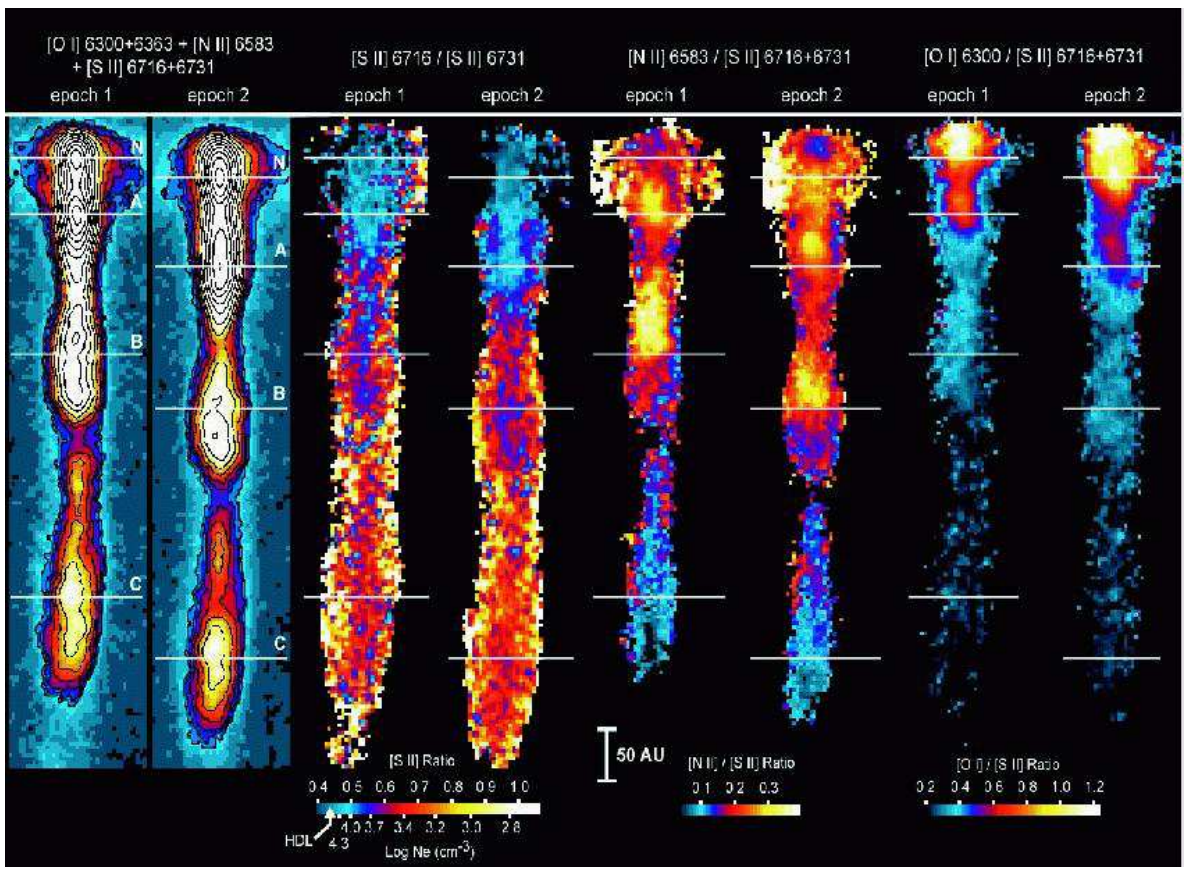}
\caption{Emission line images and ratios for epochs 1 and 2. Horizontal white lines mark
the locations of the bright knots N, A, B, and C. Each of the three emission-line ratios
has its own scale bar, color-coded for that ratio.
The [S~II] image ratio shows the jet is denser along its axis and
closer to the source. Areas of high [N~II]/[S~II] and [O~I]/[S~II]
on the side closest to the source in knots A and B move outward with the jet, and both ratios
show a gradual decline with distance from the source,
whose position is denoted by a horizontal white line.}
}
\end{figure}

\begin{figure}
\def\thefigure{10}
\vbox to 8.35in{\null\vfill\includegraphics{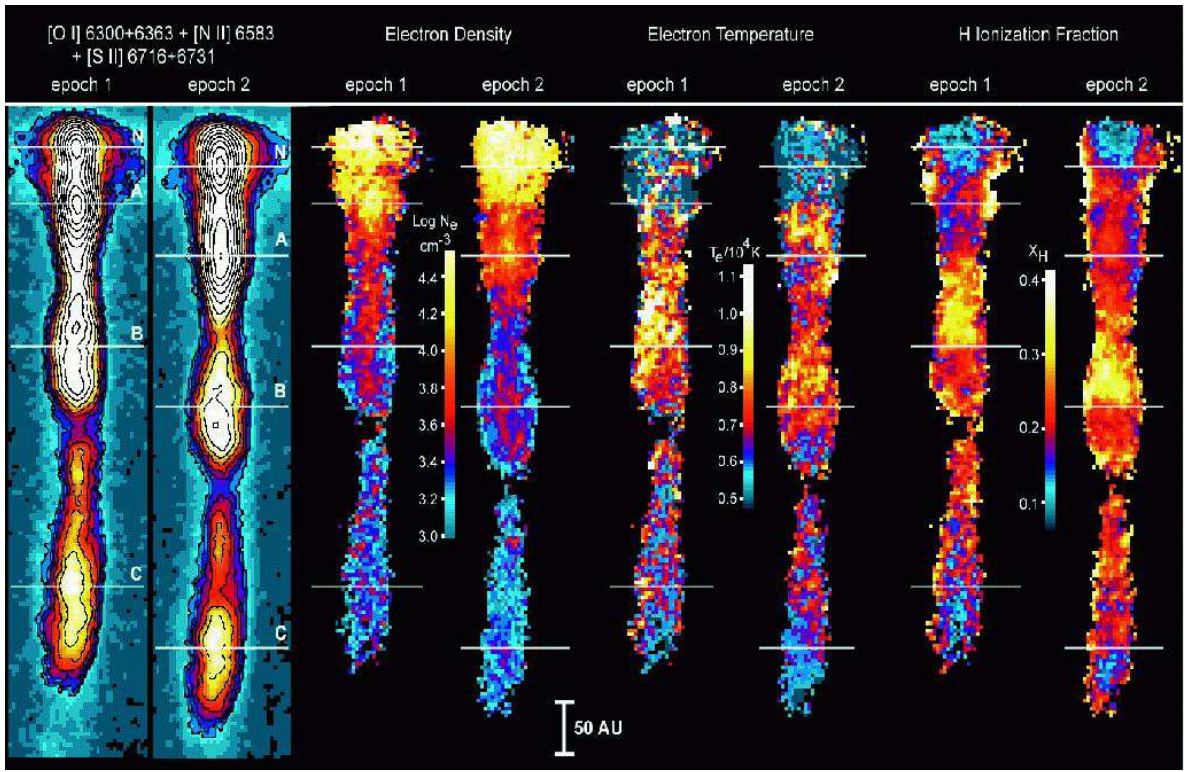}
\caption{Same as Fig.~9 but for the electron temperature, electron density and hydrogen ionization
fraction X$_H$ = H~II/(H~I+H~II). The area of high excitation (high X$_H$ and T$_e$,
color coded as yellow and white) in
knot B moves outward with the flow.  The jet emerges with a very low ($\lesssim$ 0.1)
ionization fraction, and becomes more ionized beyond $\sim$ 50~AU. At large distances
the ionization declines.
}
}
\end{figure}

\begin{figure}
\def\thefigure{11}
\vbox to 8.35in{\null\vfill\includegraphics{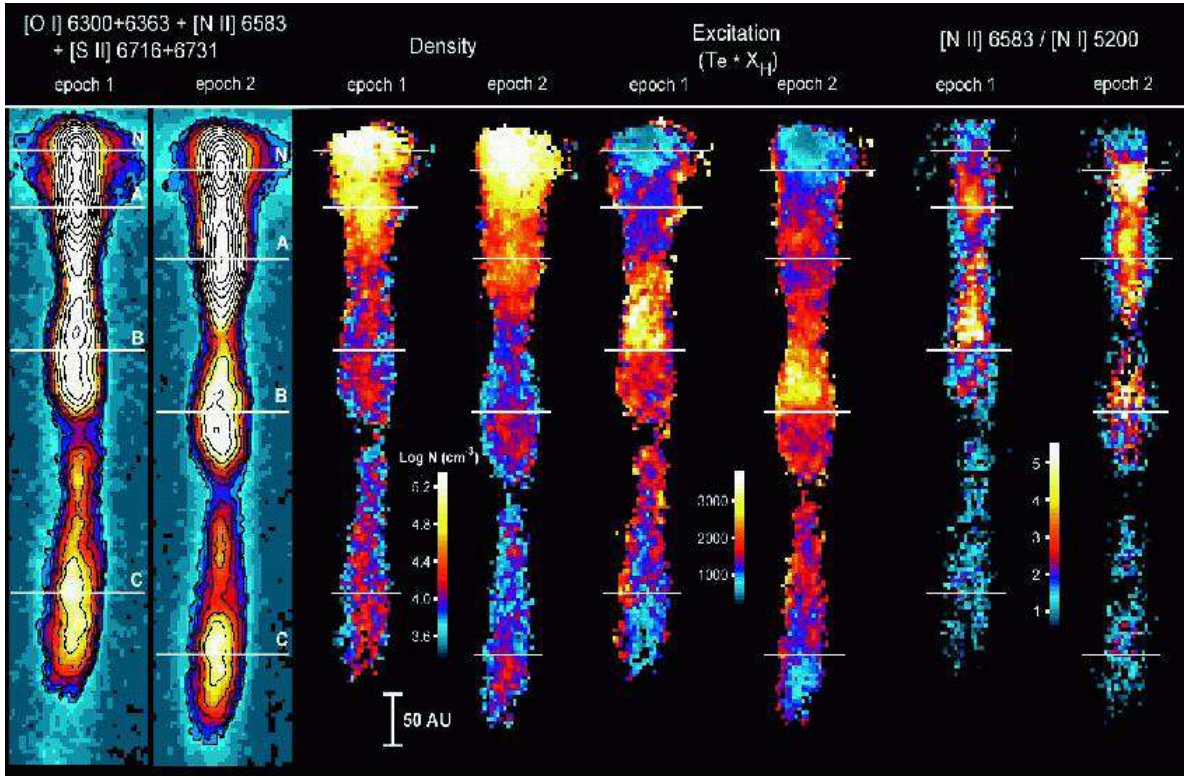}
\caption{Same as Figure 9 but for the total density, excitation, and [N~II] 6583/[N~I] 5200 ratio.
The total density rises to a value of $\sim$ $4\times 10^5$ near the source, and is higher 
(red and yellow colors) in the bright knots and along
the axis of the jet. Dense filaments, most easily seen in
the epoch 2 image, are present within knot B.  The excitation image, created by the product of
the ionization fraction and the electron temperature, shows several areas of heating between
$\sim$ 50~AU and 500~AU from the star. The [N~II]/[N~I] ratio, which is not dereddened,
declines close to the source even though the density and reddening both increase toward the source,
and both factors act to increase the ratio. The only explanation for the decline of the
ratio close to the source is that the jet is more neutral there, in agreement with the
ionization map in Figure 10.
}
}
\end{figure}

\begin{figure}
\def\thefigure{12}
\vbox to 8.35in{\null\vfill\includegraphics{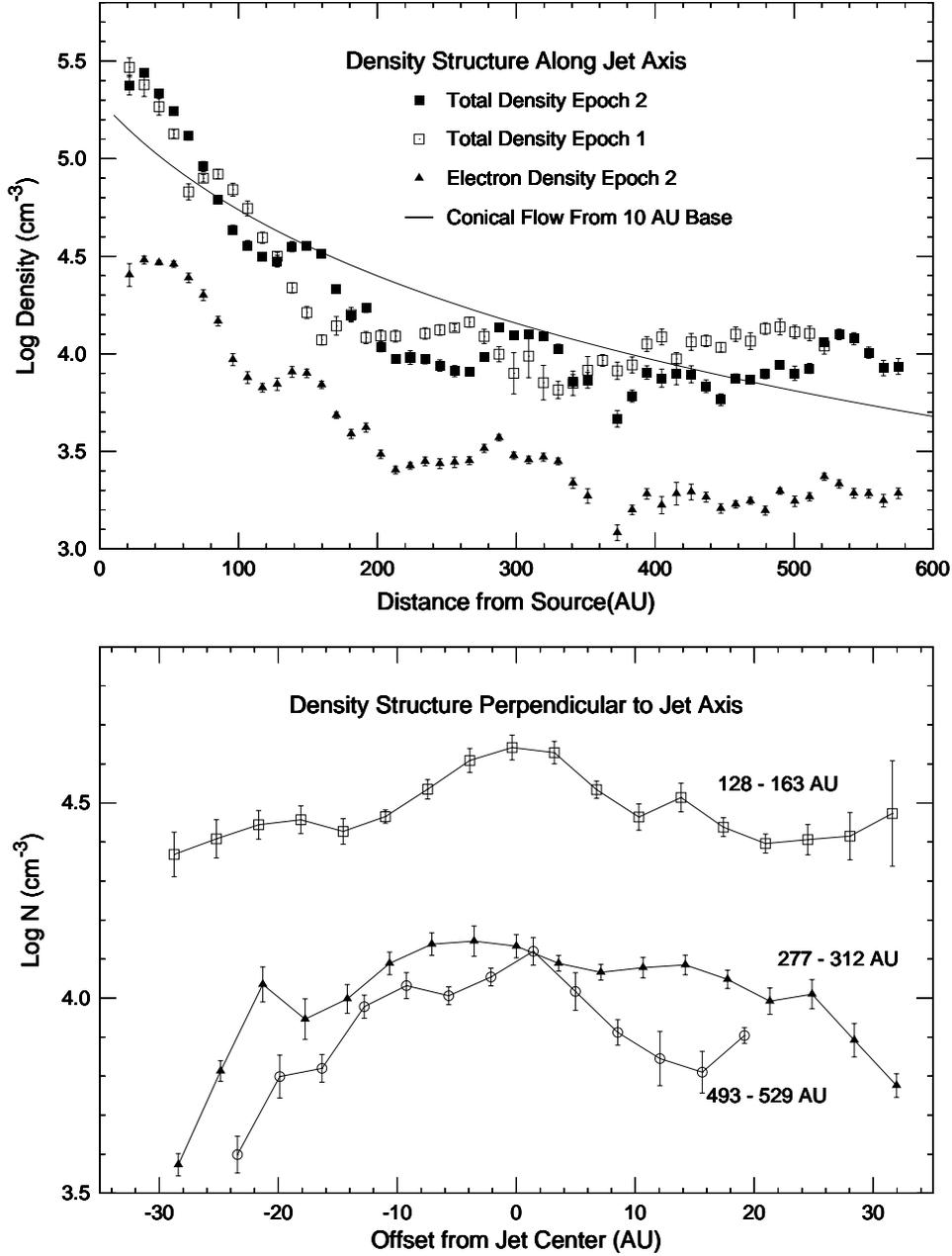}
\caption{Density structure in the HH 30 jet. Top: The electron density and the total density
decline with distance from the source. The mean and standard deviation of the electron density
and total density in a box that extends 3 pixels along the axis of the jet and 9 pixels
perpendicular to the jet produce each point and its errorbar. Density fluctuations within
the flow move along at the jet velocity between epochs. The solid curve depicts a conical
flow emerging from a base with a radius of 10~AU. Bottom: Density perpendicular to
the flow axis in epoch 2 calculated over three distance ranges from the source. The density,
calculated from the ratio of the N$_e$ and X$_H$ images,
is larger by about a factor of two on axis relative to the edges of the jet.}
}
\end{figure}

\normalsize


\clearpage

\begin{thebibliography}{}

\bibitem[Appenzeller et al.(2005)]{app05}
Appenzeller, I., Bertout, C., \& Stahl, O. 2005, A\&A 434, 1005

\bibitem[Bacciotti \& Eisl\"offel(1999)]{be99}
Bacciotti, F., \& Eisl\"offel, J.  1999, A\&A 342, 717

\bibitem[Bacciotti et al.(1999)]{ber99}
Bacciotti, F., Eisl\"offel, J., \& Ray, T.  1999, A\&A 350, 917

\bibitem[Bacciotti et al.(2000)]{bacc00}
Bacciotti, F., Mundr, R., Ray, T. Eisl\"offel, J., Solf, J.,
\& Camenzind, M. 2000, ApJ 537, L49 

\bibitem[Baluja \& Zeippen(1988)]{bz88}
Baluja, K., \& Zeippen, C. 1988, J. Phys. B. Atom. Mol. Opt. Phys. 21, 1455

\bibitem[Bell et al.(1995)]{bell95}
Bell, K., Hibbert, A., \& Stafford R. 1995, Phys. Scr. 52, 240

\bibitem[Blacher (2003)]{blacher03}
Blacher, R. 2003, Journal of Multivariate Analysis 87, 2

\bibitem[Berrington \& Burke(1981)]{bb81}
Berrington, K., \& Burke, P. 1981, Plan. Sp. Sci. 29, 377

\bibitem[B\"ohm(1956)]{bohm56}
B\"ohm, K.-H. 1956, ApJ 123, 379

\bibitem[Box(1954)]{box54}
Box, G. 1954, Ann.~Math.~Statistics 25, 290

\bibitem[Brugel et al.(1981)]{bbm81}
Brugel, E., B\"ohm, K.-H., \& Mannery, E. 1981, ApJS 47, 117

\bibitem[Butler \& Zeippen(1984)]{bz84}
Butler, K., \& Zeippen, C. 1984, A\&A 141, 274

\bibitem[Burrows et al.(1996)]{burrows96}
Burrows, C. et al. 1996, ApJ 473, 437 [B96]

\bibitem[Cochran(1934)]{cochran34}
Cochran, W. 1934, Proc.~Cambridge~Philos.~Soc. 30, 178

\bibitem[Cotera et al.(2001)]{cotera01}
Cotera, A. et al. 2001, ApJ 556, 958

\bibitem[Dopita et al.(1976)]{dmr76}
Dopita, M., Mason, D., \& Robb, W. 1976, ApJ 207, 102

\bibitem[Ferreira et al.(2006)]{ferr06}
Ferreira, J., Dougados, C., \& Cabrit, S. 2006, A\&A 453, 785

\bibitem[Froese-Fischer \& Saha (1985)]{froese85}
Froese-Fischer, C., \& Saha, H. 1985, Phys. Scr. 32, 181

\bibitem[Fruchter \& Hook(2002)]{fh02}
Fruchter, A. \& Houk, R. 2002, PASP 114, 144

\bibitem[Hartigan \& Kenyon(2003)]{hk03}
Hartigan, P., \& Kenyon, S. 2003, ApJ 583, 334

\bibitem[Hartigan et al.(2001)]{hartigan01}
Hartigan, P., Morse, J., Reipurth, B., Heathcote, S., \& Bally, J. 2001, ApJ 559, L157

\bibitem[Hartigan et al.(2004)]{hep04}
Hartigan, P., Edwards, S., \& Pierson, R. 2004, ApJ 609, 261

\bibitem[Hartigan et al.(2007)]{hartigan07}
Hartigan, P., Frank, A., Varniere, P., \& Blackman, E. 2007, ApJ in press

\bibitem[Heathcote et~al.(1996)]{heathcote96}
Heathcote, S. Morse, J., Hartigan, P., Reipurth, B., Schwartz, R.,
Bally, J., \& Stone, J. 1996, \aj\ 112, 1141 

\bibitem[Herbig (1972)]{lickbull}
Herbig, G. 1972, Lick Obs. Bulletin No.658

\bibitem[Hoog \& Craig(1995)]{hc95}
Hoog, R., \& Craig, A. 1995, ``Introduction to Mathematical Statistics'',
(New Jersey:Prentice Hill), p482

\bibitem[Hudson \& Bell(2005)]{hb05}
Hudson, C. \& Bell, K. 2005, A\&A 430, 725

\bibitem[Keenan et al.(1996)]{keenan96}
Keenan, F., Aller, L., Bell, K., Hyung, S., McKenna, F., \& Ramsbottom, C.
1996, MNRAS 281, 1073

\bibitem[Kenyon et al.(1994)]{kenyon94}
Kenyon, S., Dobrzycka, D., \& Hartmann, L. 1994, AJ 108, 1872

\bibitem[Kingdon \& Ferland(1996)]{kf96}
Kingdon, J. \& Ferland, G. 1996, ApJS 106, 205

\bibitem[Landini \& Monsignori Fossi(1990)]{lmf90}
Landini, M., \& Monsignori Fossi, B. 1990, A\&AS 82, 229

\bibitem[Le Dourneuf \& Nesbet(1976)]{ln76}
Le Dourneuf, M., \& Nesbet, R. 1976, J. Phys. B Atom. Molec. Phys. 9, L241

\bibitem[Lopez et al.(1996)]{lopez96}
Lopez, R., Riera, A., Raga, A., Anglada, G., Lopez, J., Noriega-Crespo, A., 
\& Estalella, R. 1996, MNRAS 282, 470

\bibitem[Lopez-Martin et al.(2003)]{lm03}
Lopez-Martin, L., Dougados, C., \& Cabrit, S. 2003, A\&A 405, L1 

\bibitem[McLaughlin \& Bell(1993)]{mb93}
McLaughlin, B., \& Bell, K. 1993, ApJ 408, 753

\bibitem[Mendoza (1983)]{m83}
Mendoza, C. 1983, IAU Symp. 103, {\it Planetary Nebulae}, D. R. Flower ed.,
(Dordrecht: Reidel), p143

\bibitem[Mendoza \& Zeippen(1982)]{mz82}
Mendoza, C., \& Zeippen, C. 1982, MNRAS 198, 111

\bibitem[Mundt \& Fried(1983)]{mundt83}
Mundt, R., \& Fried, J. 1983, ApJ 274, L83

\bibitem[Mundt et al.(1990)]{mundt90}
Mundt, R., Ray, T., B\"uhrke, T., Raga, A., \& Solf, J. 1990, A\&A 232, 37

\bibitem[Mundt et al.(1991)]{mundt91}
Mundt, R., Ray, T., \& Raga, A. 1991, A\&A 252, 740

\bibitem[Nussbaumer \& Storey(1983)]{ns83}
Nussbaumer, H., \& Storey, P. 1983, A\&A 126, 75

\bibitem[Pequinot et al.(1991)]{peq91}
Pequinot, D., Petitjean, P., \& Boisson, C.  1991, A\&A 251, 680

\bibitem[Pradhan(1976)]{pradhan76}
Pradhan, A. 1976, MNRAS 177, 31P

\bibitem[Podio et al.(2006)]{podio06}
Podio, L., Bacciotti, F., Nisini, B., Eisl\"offel, J., Massi, F., Giannini, T., \& Ray, T.
2006, A\&A 456, 189

\bibitem[Ray et al.(1996)]{ray96}
Ray, T., Mundt, R., Dyson, J., Falle, S., \& Raga, A. 1996, ApJ 468, L103

\bibitem[Storey \& Zeippen(2000)]{sz00}
Storey, P. \& Zeippen, C. 2000, MNRAS 312, 813

\bibitem[Verner et al.(1996)]{verner96}
Verner, D., Ferland, G., Korista, K., \& Yakovlev, D. 1996, ApJ 465, 487

\bibitem[Voronov(1997)]{voronov97}
Voronov, G. 1997, ADNDT 65, 1.

\bibitem[Wang et al.(2004)]{wang04}
Wang W., Liu, X.-W., Zhang, Y., \& Barlow, M. 2004, A\&A 427, 873

\bibitem[Watson \& Stapelfeldt(2004)]{watson04}
Watson, A., \& Stapelfeldt, K. 2004, ApJ 602, 860 

\bibitem[Wood et al.(1998)]{wood98a}
Wood, K., Kenyon, S., Whitney, B., \& Turnbull, M. 1998, ApJ 497, 404

\bibitem[Wood \& Whitney(1998)]{wood98b}
Wood, K., \& Whitney, B. 1998, ApJ 506, L43

\bibitem[Wood et al.(2000)]{wood00}
Wood, K., Wolk, S., Stanek, K., Leussis, G., Stassun, K., Wolff, M., \& Whitney, B. 2000, ApJ 542, L21

\bibitem[Wood et al.(2002)]{wood02}
Wood, K., Wolff, M., Bjorkman, J., \& Whitney, B. 2002, ApJ 564, 887

\bibitem[Zeippen(1982)]{zeip82}
Zeippen, C. 1982, MNRAS 198, 111

\bibitem[Zeippen(1987)]{zeip87}
Zeippen, C. 1987, A\&A 173, 410
\end{thebibliography}
\end{document}